\documentclass[11pt, a4paper]{article}
\pdfoutput=1
\usepackage{jheppub}
\usepackage[english]{babel}
\usepackage{amsmath,amsthm,amssymb}
\usepackage{amsfonts}
\usepackage{color}
\usepackage{graphicx}
\usepackage{url}
\usepackage{subfig}
\usepackage{enumerate}
\usepackage{pgfplots}
\usepackage{tikz}
\usepackage[section]{placeins}
\usepackage{makeidx}
\usepackage{mathtools}
\usepackage{wrapfig}
\usepackage[linewidth=1.2pt]{mdframed}
\usepackage{relsize}
\usepackage{bbm}

\def\bx{{\bf x}}

\def\bk{{\bf k}}

\def\bp{{\bf p}}

\newcommand{\om}{\omega}
\newcommand{\Om}{\Omega}
\newcommand{\ka}{\kappa}

\newcommand{\be}{\beta}

\newcommand{\al}{\alpha}

\newcommand{\si}{{\sigma}}

\newcommand{\del}{{\partial}}

\newcommand{\mc}[1]{\mathcal{#1}}

\newcommand{\sfrac}[2]{{\textstyle\frac{#1}{#2}}}
\newcommand{\half}{\sfrac{1}{2}}

\newcommand{\nn}{\nonumber}
\newcommand{\ep}{\varepsilon}

\newcommand{\gv}[1]{\ensuremath{\mbox{\boldmath $#1$}}}

\newcommand{\integral}{\int\!\! \mathrm{d}}

\newcommand{\ket}[1]{\left| #1 \right>} 

\newcommand{\wavepacket}[1]{
	\overset{
		\raisebox{-.15 ex}{
			\makebox[0 pt][c]{
				\hspace{-1 ex}
				\begin{tikzpicture}[scale=.02]
				\begin{axis}[ 
				domain=-2:2,
				no markers,
				samples=30,
				xscale=1.5,
				axis lines=none,
				] 
				\addplot[color=black,line width=20pt]
				{2.718^(-x*x)};
				\end{axis}
				\end{tikzpicture}
		}}
	}
	{#1}
}

\newcommand{\mb}{\mathbf}

\newcommand{\ketmink}{\ket{0}_{\!\ms M}}
\renewcommand{\Re}{\mathrm{Re}}

\newcommand{\csch}{\mathrm{csch}}
\let\perptmp\perp
\renewcommand{\perp}{{\! \mathsmaller{\perptmp}}}
\newcommand{\e}{\mathlarger{e}}
\newcommand{\ms}{\mathsmaller}
\newcommand{\mrm}{\mathrm}

\newcommand{\nodagger}{{\phantom{\dagger}}}
\newcommand{\mink}{{\,_{\!\!\ms{M}}}} 
\newcommand{\bdb}{\langle \hat b_k^\dagger \hat b^\nodagger_k \rangle}
\newcommand{\ddd}{\langle \hat d_k^\dagger \hat d^\nodagger_k \rangle}

\title{Entangled wavepackets in the vacuum}

\author[a,b]{Joris Kattem\"olle,}
\author[b,c]{Ben Freivogel}

\affiliation[a]{QuSoft, CWI,\\Science Park 123, Amsterdam, The Netherlands }
\affiliation[b]{Institute for Theoretical Physics, University of Amsterdam,\\Science Park 904, Amsterdam, The Netherlands}
\affiliation[c]{GRAPPA, University of Amsterdam,\\Science Park 904, Amsterdam, The Netherlands}

\emailAdd{j.j.kattemolle@uva.nl}
\emailAdd{benfreivogel@gmail.com}

\abstract{	
	Motivated by the black hole firewall problem, we find highly entangled pairs of spatially localized modes in quantum field theory. We demonstrate that appropriately chosen wavepackets localized outside the horizon are nearly purified by `mirror' modes behind the horizon. In addition, we calculate the entanglement entropy of a single localized wavepacket in the Minkowski vacuum. In all cases we study, the quantum state of the system becomes pure in the limit that the wavepackets delocalize; we quantify the trade-off between localization and purity.
}

\begin{document}
\maketitle
\flushbottom

\part{Introduction, methods and results}
\section{Introduction}
An important tool in analyzing the black hole information problem is the entanglement of particles behind the horizon with particles outside the horizon. More precisely, the quantum fields can be decomposed into modes localized outside or inside the horizon. Assuming that the quantum state approaches the Minkowski vacuum at short distances, modes outside the horizon are entangled with modes behind the horizon.

In their famous `firewall' paper \cite{AMPS}, Almheiri, Marolf, Polchinski, and Sully (AMPS) demonstrated a remarkable conflict between three fundamental principles of physics: the equivalence principle, unitarity, and causality. To accomplish this, they made use of an entangled pair of modes, which we will refer to as $H$ and $P$. These modes must have the following three properties:
\begin{itemize}
	\item Each mode separately is in a mixed state. This means their von Neumann entropies satisfy
	\begin{equation*}
	S_H \sim S_P \gtrsim 1.
	\end{equation*}
	\item The two modes purify each other, so that the combined system is nearly in a pure state.  That is
	\begin{equation*}
	S_{HP} \ll 1.
	\end{equation*}
	\item Both modes are localized in a region small compared to the Schwarzschild radius of the black hole.
\end{itemize}

To the best of our knowledge, the existence of pairs of modes satisfying these three properties has not been demonstrated.  In this paper, we  construct such entangled pairs. We analyze the problem in the simplest context of free scalar field theory in Minkowski spacetime. Because the whole point in the black hole context is to construct wavepackets which are small compared to the Schwarzschild radius, the geometry is effectively Minkowski spacetime at these distance scales. Even in this simple context, the existing literature (in particular \cite{Audretsch}) does not contain wavepackets satisfying these three criteria. 

\begin{figure}[h]
	\centering
	\def\svgwidth{1\textwidth}
	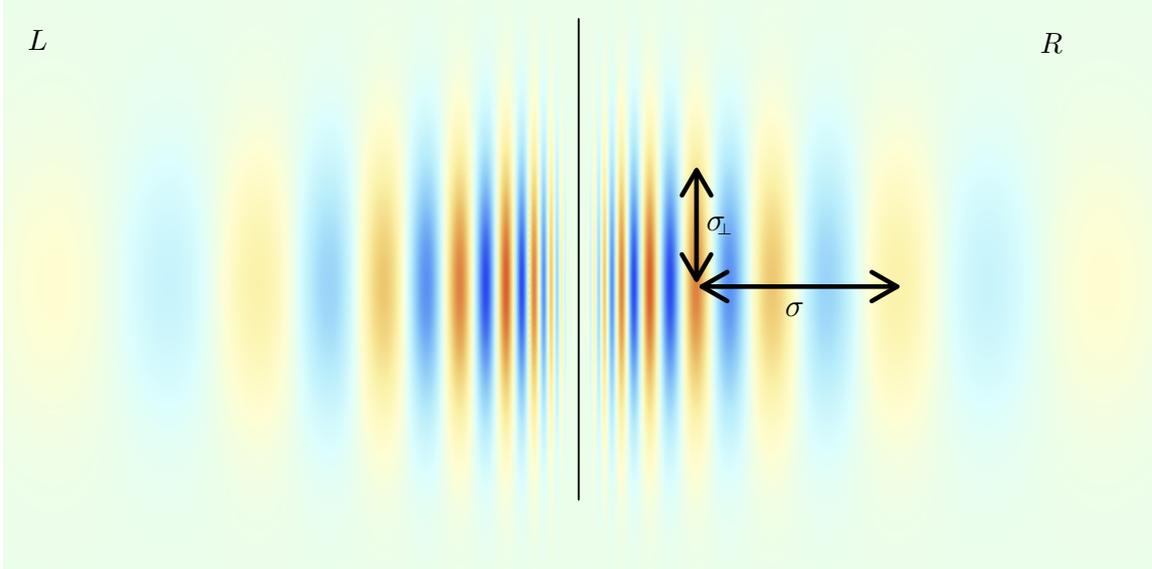\caption{Real part of a Rindler wavepacket and its `mirror' mode. The parameters are set such that the depiction is most clear, and not to make the wavepackets actually highly entangled. Conditions for the wavepackets to be highly entangled are derived in this paper. (For later reference: $\si_\perp=\si, k=4\pi/\si$.)  \label{fig:pairOfRindlerWavepackets}}
\end{figure}

Modes satisfying the first two criteria are well known: they are the `Rindler modes', which we refer to as Rindler plane waves. These modes, however, are not localized. The wavepackets we consider are simply Rindler plane waves modulated with a Gaussian envelope. They are most naturally written in terms of the Rindler coordinate $\xi$, which is related to the proper distance from the Rindler horizon by
\begin{equation*}
{\rm distance} \sim e^\xi.
\end{equation*}
In terms of this coordinate, our modes are plane waves with Gaussian envelopes of length $\si$ (see figure~\ref{fig:pairOfRindlerWavepackets}).

We find that for appropriately chosen parameters, such a wavepacket localized to the right of the Rindler horizon is nearly purified by a corresponding `mirror' wavepacket to the left of the horizon.

Quantitatively, we find 
\begin{itemize}
	\item There is a trade-off between localization and purification. In 1+1 dimensions, the entropy of the combined system is
	\begin{equation}
	\label{eq:introSHP}
	S_{HP} \sim {\log(\sigma) \over \sigma^2}.
	\end{equation}
	\item In higher dimensions, the proper width $\si_\perp$ of the wavepacket  in the directions  along the horizon must be large compared to the radial length $\sigma$ in order for the combined system to behave approximately like a pair of Rindler plane waves,
	\begin{equation}
	\si_\perp > \frac{1}{2} \e^{\si^2}.
	\end{equation}
\end{itemize}
If the latter condition is not satisfied, the difference is surprisingly large and the transition seems to behave as a phase transition. We know of no simple physical argument for this.

Our results quantify the extent to which the degrees of freedom of quantum field theory can be organized into localized, entangled pairs. The existence of such highly entangled pairs puts the `firewall' argument of AMPS on stronger footing. In addition, the quantitative results for how localized the wavepackets can be while remaining in a pure state gives an indication of on which scales physics must be modified in order to avoid the paradox.

We also calculate the entanglement entropy of a single Gaussian wavepacket, which is approximately a momentum eigenstate with momentum $k$ and has length $\si$ and width $\si_\perp$. In this case we find that for large spatial extension the entropy goes as a power law in $1/\si_\perp$,
\begin{equation}
S \sim \frac{1}{(k \si_\perp)^4},
\end{equation}
where we have neglected logarithmic terms.
For infinite width, and in $1+1$ dimensions, the entropy is exponentially small in the length $\si$ of the wavepacket,
\begin{equation}
\label{eq:introS}
S \sim e^{-(k \sigma)^2},
\end{equation}
where we have kept only the exponential dependence.
Surprisingly, in the limit that the width goes to infinity but the length to zero (i.e. $\si_\perp\to\infty$, $\si\to 0$), the entanglement entropy remains finite, approaching the value $S \approx 0.35\,$bit. 
The latter also holds for our Rindler wavepacket in the same limit. 

A number of interesting open questions remain. 
\begin{itemize}
\item Does including interactions, such as gravitational backreaction, change our conclusions?
\item We have considered modes with a Gaussian envelope. How does the entanglement change for modes that are truly localized in a region of space?
\item Is there a better choice of modes that allows for a purer state with the same localization? We have not shown that our modes are optimal in this respect. It would be very interesting to know how universal our results quoted above are (eqs. \ref{eq:introSHP}-\ref{eq:introS}).
\item{What physical arguments explain the sudden transition at $\si_\perp\approx\half e^{\si^2}$, and the entropy of $S\approx0.35\,$bit in the limit $\si_\perp\to\infty$, $\si\to 0$?}
\end{itemize}

This paper is structured as follows. It is divided into two parts. The current, first part contains some preliminaries (section \ref{sec:preliminaries}), a general recipe for computing the entropy of a collection of modes in the Minkowski vacuum (section \ref{sec:theEntanglementEntropyOfASetOfModesInTheMinkowskiVacuum}),
the usage of this recipe to compute the entropy of various wavepackets (sections \ref{sec:MinkowskiWavepacket} and \ref{sec:aPairOfRindlerWavepackets}), a discussion on how these results apply to the firewall paradox (section  \ref{sec:applicationToTheFirewallParadox}) and a conclusion (section \ref{sec:conclusion}). The majority of calculations can be found in part II of this paper. 

{\bf Note:} This work originated as the master's thesis of the first author. A more detailed account of the background and techniques can be found in the thesis \cite{Kattemolle}.
\section{Preliminaries}\label{sec:preliminaries}
Here we give a short introduction to the concept of particles in the vacuum, Gaussian states, and Rindler space. Useful existing literature on these topics includes references \cite{BirellDavies, Carroll, Ferraro, Adesso, Demarie, Wang, Takagi, Crispino, Susskind}. We set $c=\hbar=k_B=1$. 

\subsection{Particles in the vacuum}
\label{sec:particlesInTheVacuum}
Consider the free massless scalar field. The field operator $\hat \phi$ can be expanded over the basis of Minkowski plane waves $f_\bp$ as
\begin{align*}
	\hat \phi=\integral \bp\,(f_\bp \hat a_\bp+f_\bp^* \hat a_\bp^\dagger),
\end{align*}
thus associating the mode operator $\hat a_\bp$ with the Minkowski plane wave $f_\bp$. The Minkowski vacuum $\ket{0}_\mink$ is defined as the state for which
\begin{align}
\label{eq:MinkowskiVacuum}
\hat a_\bp\ketmink=0\qquad \text{for all }\bp.  
\end{align}

Alternatively, the field could be expanded over a basis $\{g_\bk\}$,
\begin{align*}
	\hat \phi=\integral \bk\,(g_\bk \hat b_\bk+g_\bk^* \hat b_\bk^\dagger).
\end{align*}
Here $\bk$ need not be momentum, but should be thought of as an index that labels the basis modes. Thus the operator $\hat b_\bk$ is associated with the mode $g_\bk$. This operator can be written in terms of the mode operators $\hat a_\bp$ as 
\begin{align}
\label{eq:bInTermsOfa}
\hat b_\bk= \integral \bp \left(\al_{\bk\bp}^* \hat a_\bp-\be_{\bk\bp}^*\hat a_{\bp}^\dagger\right),
\end{align}
where $\al_{\bk\bp}$ and $\be_{\bk\bp}$  are the so-called Bogolyubov coefficients. These coefficients can be computed by
\begin{align}
\label{eq:alphaBeta}
\al_{\bk\bp}=(g_{\bk},f_{\bp}),
&&
\be_{\bk\bp}=-(g_{\bk},f_{\bp}^*),
\end{align} 
where ${}^*$ is complex conjugation, and the inner product $(\cdot,\cdot)$ is given by the Klein-Gordon inner product. In Minkowski space it reads 
\begin{align}
\label{eq:KleinGordonInnerProduct}
(g,f)=-i \integral\bx \left[\,g\,\del_t f^*-(\del_t g)f^*\right].
\end{align}

Now by using (\ref{eq:MinkowskiVacuum}) and (\ref{eq:bInTermsOfa}), we see that the expectation value of the number operator of the new modes equals
\begin{align}
\label{eq:particlesInTheVacuum}\langle \hat b_\bk^\dagger \hat b^\nodagger_\bk \rangle_\mink &=\integral \bp\,|\be_{\bk\bp}|^2.
\end{align}
So, even though the overall state is the vacuum with respect to the modes $f_\bp$, the modes $g_\bk$ are in an excited state. As will become clear in due course, the modes $g_\bk$ will \emph{separately} often be in a mixed state due to entanglement with other modes. Thus they will have some non-zero von Neumann entropy. Since the overall state $\ket{0}_\mink$ is pure, this entropy is solely due to entanglement, and we will therefore refer to this quantity as \emph{the entanglement entropy of the mode}. In this paper, the overall state is always the Minkowski vacuum, so we will henceforth drop the subscript $M$. 

\subsection{Gaussian states}\label{sec:GaussianStates}
The Minkowski vacuum is in fact a \emph{Gaussian state} (see 
\cite{Wang, Kattemolle} for a proof.) We will here show how Gaussian states and continuous variable quantum information theory can be used to calculate various quantities. More detail can be found in \cite{Ferraro, Adesso, Demarie, Wang}. 

In the phase space formulation of quantum mechanics, any state~-~be it mixed or pure~-~is fully described by the characteristic function $\chi$. There is a specific one-to-one relation between the density matrices and the characteristic functions \cite{Adesso}, but its actual form is not of interest here. 

Now let $\mb{X}=(q_1,p_1,\ldots,q_{\ms N},p_{\ms N} )^{\ms T}$ be a vector in the $2N$-dimensional phase space of an $N$-mode system with mode operators $\hat b_m$. A Gaussian state then, is a state with a Gaussian characteristic function, 
\begin{align}
\label{eq:characteristicFunction}
\chi(\ms\bx)\sim e^{-\half  \bx^T \gv \si \bx}.
\end{align}
Here $\gv \sigma$ is the covariance matrix,
\begin{align}
\label{eq:covarianceMatrix}
	[\gv \si]_{kp}=\half\langle\{\hat R_k,\hat R_p\}\rangle-\langle \hat R_k\rangle\langle \hat R_p\rangle,
\end{align}
with $\{\cdot,\cdot\}$ the anti-commutator, $
\hat{\mb{R}}=(\hat q_1,\hat p_1, \ldots,\hat q_{\mathsmaller N},\hat p_{\mathsmaller N})^T$,
and $(\hat q_m,\hat p_m)$ the quadrature operators
\begin{align}
\label{eq:quadratures}
\hat q_m=\frac{1}{\sqrt{2}}(\hat b_m+\hat b_m^\dagger), &&\hat p_m=\frac{1}{i\sqrt{2}}(\hat b_m-\hat b_m^\dagger).
\end{align}
Note that in the standard formulation of quantum mechanics, the density matrix is already infinity dimensional for $N=1$, whereas the covariance  matrix is only $2N$-dimensional in general. 

Since $\gv \si$ contains all information about the Gaussian state it must, in principle, be possible to express the von Neumann entropy $S$ of that state in terms of the entries of $\gv \si$ only. Indeed there is a nice formula that does exactly that \cite{Demarie}. It reads
\begin{align}
\label{eq:entropy}	
S=\sum_{j} \left[\left(s_j+\half\right)\log\left(s_j+\half\right)-\left(s_j-\half\right)\log\left(s_j-\half\right)\right],
\end{align}
where the $s_j$ are the positive eigenvalues of the matrix $i\gv\Om\gv\si$, called the \emph{symplectic eigenvalues} of $\gv\si$, and 
\begin{align*}	
\gv\Omega=\bigoplus_{j=1}^{\mathsmaller N}
\begin{pmatrix*}[r]
0 & \ 1\ \\
-1 & \ 0\ \\
\end{pmatrix*}.
\end{align*}
\paragraph{Example}
Let us write out the entropy of a single mode. For $N=1$, the covariance matrix reads
\begin{align*}
\gv \sigma =
\begin{pmatrix}
	\langle \hat q^2\rangle & \half\langle \{\hat q,\hat p\}\rangle\\
	 \half\langle \{\hat q,\hat p\}\rangle &\langle \hat  p^2\rangle\\
\end{pmatrix}.
\end{align*} 
It has a single symplectic eigenvalue, $s=\sqrt{\langle \hat p^2\rangle\langle \hat q^2\rangle-\langle \{\hat q,\hat p\}\rangle^2}$.

It is sometimes more convenient to have this symplectic eigenvalue in terms of the mode operator $\hat b_m$ rather than the quadrature operators. Using (\ref{eq:quadratures}) and the bosonic commutation relations, one finds
\begin{align}
\label{eq:symplecticEigenvalueOfOneMode}
	s=\sqrt{\left(\langle \hat b_k^\dagger \hat b^\nodagger_k \rangle + \half \right)^2-|\langle \hat b_k \hat b_k\rangle|^2}\ .
\end{align}
Plugging this into  (\ref{eq:entropy}) directly yields the entropy in terms of the expectation values. In the cases that follow, we will find the closed form of these expectation values by using (\ref{eq:bInTermsOfa}) and (\ref{eq:KleinGordonInnerProduct}).

\subsection{Rindler space}
\label{sec:RindlerSpace}

Rindler space is Minkowski space as seen by an accelerating observer. In effect it is a mere coordinate transformation to coordinates that are most natural for the accelerating observer. For a more detailed description than the following, see e.g. \cite{BirellDavies, Carroll}.

The transformation from Minkowski coordinates $(t,x)$ to Rindler coordinates $(\eta,\xi)$ comes in four patches or `wedges'. The first, $x>|t|$, is called the right Rindler wedge $R$. Here, the transformation is defined as
\begin{align*}
t=\frac{1}{a}\,e^{a\,\xi^\mathsmaller R}\sinh(a\eta^\mathsmaller R),&&x=\frac{1}{a}\,e^{a\,\xi^\mathsmaller R}\cosh(a\eta^\mathsmaller R),
\end{align*}
where $a$ sets the energy scale. We have set $a=1$ in the introduction, and we will continue to do so in the following sections,  with the exception of section \ref{sec:applicationToTheFirewallParadox}.

A second region is $x<|t|$ and is called the left Rindler wedge $L$. Here the transformation is defined as
\begin{align*}
t=-\frac{1}{a}\,e^{a\,\xi^\mathsmaller L}\sinh(a\eta^\mathsmaller L),&&x=-\frac{1}{a}\,e^{a\,\xi^\mathsmaller L}\cosh(a\eta^\mathsmaller L).
\end{align*}
In terms of the Rindler coordinates the Minkowski metric $\mathrm d s^2=-\mathrm d t^2 + \mathrm d x^2$ reads
\begin{align}
\label{eq:RindlerMetirc}
	\mrm ds^2=e^{2a\xi}(-\mrm d\eta^2+\mrm d\xi^2),
\end{align}
be it in $R$ or $L$. Indeed, these coordinates are the most natural coordinates for an accelerating observer: objects standing still in the Rindler frame at $\xi=\xi_0$ have a proper acceleration of $a e^{-a\xi_0}$, and there is a horizon at $x=|t|$. Therefore the Rindler coordinates can be used to describe spacetime just outside a Schwarzschild black hole. (A more careful treatment, as for example in \cite{Susskind}, shows that indeed the Rindler metric is a good approximation to the Schwarzschild metric when the proper distance to the horizon is small enough.)

The Rindler plane waves 
\begin{align}
\label{eq:RindlerPlaneWaves}g^{\ms R}_p(\eta^{\ms R},\xi^{\ms R})=\left\{
\begin{matrix}
	\sfrac{1}{\sqrt{4\pi \om_p}}e^{i p\xi^\mathsmaller R-i \om_p \eta^\mathsmaller R}&\  &\text{in }R\\
	0&\ &\text{in } L\\
\end{matrix}
\right.,
&\ &
g^{\ms L}_p(\eta^{\ms L},\xi^{\ms L})=\left\{
\begin{matrix}
	0&\  &\text{in }R\\
	\sfrac{1}{\sqrt{4\pi \om_p}}e^{ip\xi^\mathsmaller L+i\om_p\eta^\mathsmaller L}&\ &\text{in } L\\
\end{matrix}
\right.,\end{align}
where $\om_p=|p|$,
are solutions to the equation of motion of the free, massless scalar field. They can be used as a basis to expand the field operator over, that is,  
\begin{align*}
\hat \phi =\integral p \left[\left(g^{\ms R}_p\,\hat b^{\ms R}_p+g^{\ms L}_p\,\hat b^{\ms L}_p\right)+\mathrm{h.c.}\right],
\end{align*}
where h.c. stands for the Hermitian conjugate of the preceding term.

The Rindler plane waves do not annihilate the Minkowski vacuum, that is, $\hat b^{\ms{L,R}}_k\ket{0}_\mink \neq 0$. Instead, they have their own vacuum $\ket{0}_{\! \ms R}$ defined by $\hat b^{\ms{L,R}}_k\ket{0}_{\!\ms R}=0$. It is, however, possible to express them  in terms of operators $(\hat c_p^\ms{I},\hat c_p^\ms{II})$ that do annihilate the vacuum \cite{Carroll},
\begin{equation}
\begin{aligned}
	\hat b^{\ms R}_p&=\sqrt{\half \csch(\pi \om_p / a)}\left(\e^{\pi\om_p/(2a)}\hat c_p^{\ms{I}}+\e^{-\pi\om_p/(2a)}\hat c^{\ms{II}\dagger}_{-p}\right),\\
	\hat b^{\ms{L}}_p&=\sqrt{\half \csch(\pi \om_p / a)}\left(e^{\pi\om_p/(2a)}\hat c^{\ms{II}}_p+\e^{-\pi\om_p/(2a)}\hat c^{\ms{I}\dagger}_{-p}\right).
    \end{aligned}
\end{equation}

\paragraph{Example}
We can now continue the example in section \ref{sec:GaussianStates} and calculate the entropy of a Rindler plane wave by calculating the two relevant expectation values. Using (\ref{eq:particlesInTheVacuum}), we find
\begin{align}
\label{eq:thermalSpectrum}
\langle \hat b^{\ms R}_p\hat b^{\ms R}_p\rangle =0,\qquad
\langle \hat b_p^{\ms R \dagger}\hat b^{\ms R\nodagger}_p\!\!\rangle  =\frac{1}{\e^{\frac{2\pi}{a} \om_p }-1}\equiv N_p.
\end{align}
Thus, the Rindler plane waves have a thermal spectrum $N_p$ with temperature $a/(2\pi)$. By (\ref{eq:symplecticEigenvalueOfOneMode}) the symplectic eigenvalue equals
\begin{align}
\label{eq:symplectiEigenvalueOfRindlerPlaneWave}
s=\frac{1}{\e^{\frac{2\pi}{a} \om_p }-1}+\frac{1}{2}.
\end{align}
Thus the entropy of a Rindler plane wave with Rindler momentum $p$, in $L$ or $R$, equals
\begin{align}
\label{eq:entropyOfRindlerPlaneWave}
\widetilde S_L=\widetilde S_R\equiv \widetilde S=\left(s+\half\right)\log\left(s+\half\right)-\left(s-\half\right)\log\left(s-\half\right),
\end{align}
with $s$ as in (\ref{eq:symplectiEigenvalueOfRindlerPlaneWave}).  Throughout, we will use tildes to refer to plane waves. 

In addition to a single mode, we can compute the entropy $\widetilde S_{LR}$ of the two-mode system that is comprised of a Rindler plane wave $g_k^{\ms R}$ and its `mirror' or `partner' mode $g_{-k}^{\ms L}$. We call this system a \emph{pair} of Rindler plane waves. This computation is done as in the single mode case, the only difference being the dimension of the covariance matrix. One finds that there are two symplectic eigenvalues, which are both equal to $s_{1,2}=1/2$. Plugging these into (\ref{eq:entropy}) yields 
\begin{align}
\label{eq:entropyOfAPairOfRindlerPlaneWaves}
\widetilde S_{LR}=0.
\end{align}
Thus a Rindler mode is exactly purified by its mirror mode. 

In the Minkowski vacuum all Rindler plane waves have a thermal density matrix, but are purified by their partners. Thus it can be derived that the Minkowski vacuum can be written as \cite{Harlow, Kattemolle}
\begin{align}
	\label{eq:minkowskiVacuumInTermsOfRindlerModes}
\ketmink=\bigotimes_p\left(\sqrt{{1-e^{-2\pi\om_p/a}}}\ \sum_{n=0}^\infty e^{-\frac{\pi }{a}\om_p n}\ket{n}_L\ket{n}_R\right).	
\end{align}
Here, $\ket{n}_R$ is obtained by acting with $\hat b_p^{\dagger\ms R}$ on the Rindler vacuum $n$ times, and  $\ket{n}_L$ is obtained by acting with $\hat b_{-p}^{\dagger \ms L}$ on the Rindler vacuum $n$ times. This expression shows \emph{how} the Rindler modes are entangled, and not just by how much. In this paper we investigate exactly what happens to this entanglement structure when the Rindler plane waves are localized.

\paragraph{Higher dimensions} So far, we have considered Rindler space with one temporal direction and one spatial direction. The Rindler metric (\ref{eq:RindlerMetirc}) can be extended to higher dimensions by adding a flat, $n$-dimensional hyperplane perpendicular to $\rm d \eta $ and $\rm d\xi$. The metric then reads
\begin{align*}
	\mrm ds^2=e^{2a\xi}(-\mrm d\eta^2+\mrm d\xi^2)+\mrm d\bx_\perp^2.
\end{align*}
So in total, we now have $D=1+1+n$ dimensions, where $n$ is the number of perpendicular directions. The Rindler plane waves also extend to $D$ dimensions (see eq. \ref{eq:initialValueConditionsOfDDimensionalRindlerPlaneWaves}). 

For reference later on, we note that the operators associated with these plane waves can also we written in terms of operators that annihilate the vacuum \cite{Takagi, Crispino}, 
\begin{align}
\label{eq:rindlerPlaneWavesInTermsOfModesThatAnnihilateTheVacuum }
\begin{aligned}
\hat b^{\ms R}_\bp&=\hat b^{\ms R}_{(\ms\Om,\bp\perp)}=\sqrt{\half \csch(\pi \Om)}\left(\e^{\pi\Om/2}\hat c_{(-\ms\Om,\bp_\perp)}+\e^{-\pi\Om/2}\hat c^{\dagger}_{(\ms\Om,-\bp_\perp)}\right),\\
	\hat b^{\ms{L}}_\bp&=\hat b^{\ms{L}}_{(\ms\Om,\bp\perp)}=\sqrt{\half \csch(\pi \Om)}\left(e^{\pi\Om/2}\hat c_{(\ms\Om,\bp\perp)}+\e^{-\pi\Om/2}\hat c^{\dagger}_{-(\ms\Om,\bp\perp)}\right),
	\end{aligned}
\end{align}
where the operators $\hat c$ satisfy the commutation relations
\begin{align*}
	\left[\hat c_{(\pm \ms\Om,\bp_\perp)},	\hat c_{(\pm \ms\Om',\bp_\perp')}^\dagger\right]=\delta(\Om-\Om')\delta(\bp_\perp-\bp_\perp').  
\end{align*}
 
\section{The entanglement entropy of a set of modes in the Minkowski vacuum}
\label{sec:theEntanglementEntropyOfASetOfModesInTheMinkowskiVacuum}
In the previous section we have seen that after a basis transformation, the modes in the new basis can be in an excited state even though the overall state is the Minkowski vacuum (see eq.~\ref{eq:particlesInTheVacuum}). Also, we have seen how the entropy of a general Gaussian state can be computed (eq.~\ref{eq:entropy}). As an example, we calculated the entropy of a single Rindler plane wave and a pair of Rindler plane waves using the formalism of continuous variable quantum information (eq.~\ref{eq:entropyOfRindlerPlaneWave}). We will now extend this to the most general case and give a recipe for computing the entanglement entropy of a set of $N$ modes that are mutually orthogonal under the Klein-Gordon inner product (\ref{eq:KleinGordonInnerProduct}). 

As mentioned before, the Minkowski vacuum is a Gaussian state. A Bogolyubov transformation, as described in section \ref{sec:particlesInTheVacuum}, is a mere basis transformation, so naturally the Minkowski vacuum is still a Gaussian state after such a transformation. 

Given any set of $N$ modes that are mutually orthogonal under the Klein-Gordon inner product, there exists some Bogolyubov transformation that transforms the Minkowski plane wave basis to a basis of which these $N$ modes form a subset. In principle the characteristic function (\ref{eq:characteristicFunction}) of the $N$ modes can be obtained by explicitly finding such a transformation, and then integrating out all complementary basis modes. The resulting characteristic function will still be Gaussian, and therefore equation \ref{eq:entropy} can be evoked to compute the entropy of the $N$ modes together. 

The above procedure, however, is not tractable in many cases and furthermore unnecessary. Namely, one can start from the fact that the $N$ modes are in \emph{some} Gaussian state and then compute the entries of the covariance matrix. To do so, we only need to know the Bogolyubov coefficients for some subset of indices. 

Another crucial insight that makes the calculations tractable is that we do not need to know the full time evolution of the $N$ modes we are interested in. This is because the symplectic eigenvalues only depend on some expectation values (e.g.  eq.~\ref{eq:symplecticEigenvalueOfOneMode}), which in turn only depend on some Bogolyubov coefficients (e.g. eq.~\ref{eq:particlesInTheVacuum}). These coefficients are given by some Klein-Gordon inner product (\ref{eq:alphaBeta}), which only depends on the mode functions at a given time slice and their time derivatives at that same slice. 

So to summarize: \emph{the entropy of a set of mutually orthogonal modes is a functional of the initial value conditions of these modes only.} This allows us to compute the entanglement entropy of a set of modes without even having to solve the field's equation of motion, and without having to explicitly trace out other modes. The recipe is as follows

\begin{enumerate}
\item Define the $N$ modes by their initial value conditions.

\item Using (\ref{eq:alphaBeta}), compute the necessary Bogolyubov coefficients.
\item Find out which expectation values are relevant by computing the $N$ symplectic eigenvalues in terms of expectations values (\ref{eq:entropy}).
\item Compute these expectation values using (\ref{eq:bInTermsOfa}) and the results of step 2.
\item Plug the symplectic eigenvalues, which are now in closed form, into the expression for the entropy (\ref{eq:entropy}).  
\end{enumerate}
We will follow these steps in this order in the following sections for various choices of the $N$ modes.
  
\section{Minkowski wavepacket}\label{sec:MinkowskiWavepacket}
Here we will compute the entropy of a Minkowski plane wave with a Gaussian envelope. So in this section, $N=1$. First this is done in 1+1 dimensions and then in $D$ dimensions. For the sake of example, we will be relatively explicit in the first subsection, but after that we will just state the results. Full detail on the calculations can be found in section \ref{sec:calcMinkowski}.
\subsection{1+1 dimensions}
\begin{figure}[h]
\centering
\def\svgwidth{1\textwidth}
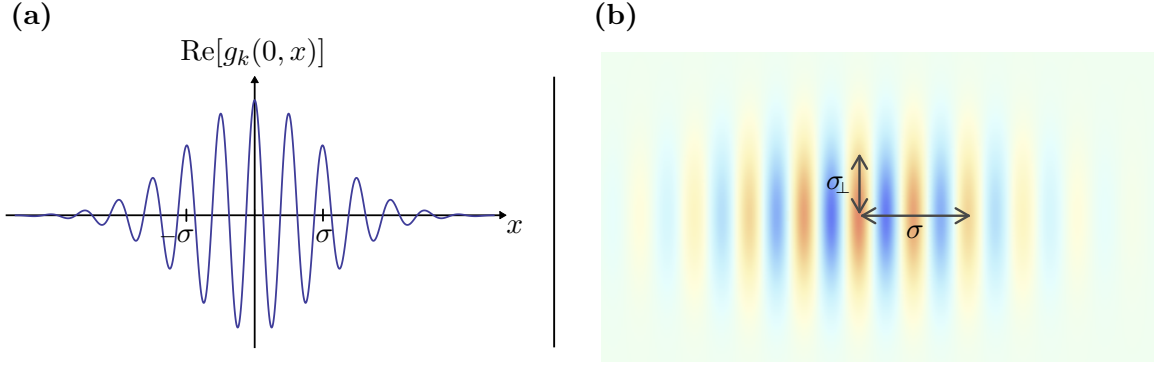\caption{Schematic depiction of: the real part of \textbf{(a)} the 1+1 dimensional Minkowski wavepacket (arbitrary units on the vertical axis), and \textbf{(b)} a slice of the $D$-dimensional Minkowski wavepacket. For the plots, we have set $k=4\pi/\si$ and $\si_\perp=\si/2$. \label{fig:MinkowskiWavepacket}}
\end{figure}

Consider the mode function
\begin{align}
\label{eq:MinkowskiWavepacket}
g_{k}(t,x)=\frac{1}{\mc N}\,\e^{-\frac{1}{2\si^ 2}(x-t)^2}\e^{ikx-i k t}.
\end{align}
Here, $\mc N$ is some normalization, $k>0$ is the approximate momentum, and $\si$ is the length of the Gaussian envelope. Note that $k$ and $\si$ should be thought of as being fixed. We can vary those parameters, but we should keep in mind that as we do so, we keep going to different bases. Figure \ref{fig:MinkowskiWavepacket}(a) shows a plot of the mode function.

This mode is actually a valid solution of the field's equation of motion, but since we know the full time evolution anyway, it is given here. Also, it is important to note that $k$ should be seen as a fixed quantity, and not as an index that labels modes in an orthogonal basis.

Now that we have defined the wavepacket, we can move on to the Bogolyubov coefficients. By solving the integrals (\ref{eq:alphaBeta}), we obtain 
\begin{align*}
\al_{kp}&=\left\{\begin{array}{ll}0&:p\leq0\\\sqrt{\frac{\si p}{ k }}\pi^{-1/4}\exp[-\frac{\si^2}{2}(k-p)^2]&:p>0\end{array}\right.,
\\
\be_{kp}&=\left\{\begin{array}{ll}0&:p\leq0\\\sqrt{\frac{\si p}{ k }}\pi^{-1/4}\exp[-\frac{\si^2}{2}(k+p)^2]&:p>0\end{array}\right..
\end{align*}
We already know what expectation values to compute (see equation \ref{eq:symplecticEigenvalueOfOneMode}).
Using (\ref{eq:bInTermsOfa}), we find
\begin{align*}
\langle\hat b_k^\dagger \hat b^\nodagger_k \rangle =\frac{1}{2} \left(\frac{e^{-\sigma ^2 k^2 }}{\sqrt{\pi } k \sigma }-\text{erfc}(k \sigma )\right),&&
\langle\hat b_k \hat b_k\rangle =-\frac{e^{-\sigma ^2 k^2}}{2 \sqrt{\pi }k\si}.
\end{align*}
Then by (\ref{eq:symplecticEigenvalueOfOneMode}), 
\begin{align*}
s=\frac{1}{2\pi^{1/4}}\sqrt{\text{erf}(k\sigma) \left(\sqrt{\pi } \text{erf}(k\sigma)+\frac{2 e^{-\sigma^2k^2}}{k\sigma}\right)}.
\end{align*}
Plugging this into the formula for the entropy (\ref{eq:entropy}) gives the entropy as a function of $k\si$. 

Let us analyze the result. First of all, note that in the limit $k\si\to 0$, the symplectic eigenvalue equals $s=1/\sqrt{\pi}$, and so
\begin{align}
\label{eq:misteryS}
	\lim_{k\si\rightarrow 0} S=S|_{s=1/\sqrt{\pi}}\approx0.35\text{ bit}.
\end{align}
Contrastingly, $\langle\hat b_k^\dagger \hat b^\nodagger_k \rangle $ goes to infinity in the same limit. In other words, an infinitely narrow wavepacket is infinitely excited, but has a finite entropy. 

Secondly, as we now increase the width of the wavepacket, $s$ is exponentially close to 1/2. Examining formula \ref{eq:entropy} for the entropy, we see that for $s = 1/2 + \varepsilon$, the entropy is of order $S \sim \varepsilon \log(1/\varepsilon)$ for small $\varepsilon$. In this case $\varepsilon \sim \exp[-(k \sigma)^2]$ (neglecting a power law prefactor), so  $S$ drops to zero exponentially fast, $S \sim \exp[-(k \sigma)^2]$, where we again keep track of only the exponential dependence. So for all intents and purposes, the localization entropy of a Gaussian Minkowski wavepacket is negligible.

\subsection{$D$ dimensions}
Before we define our $D$-dimensional Gaussian Minkowski wavepacket, let us introduce some notation. We divide the spatial vector $\bx$ into a direction \emph{parallel} to the direction the wavepacket is moving in, and the $n$ directions \emph{perpendicular} thereto. That is, $\bx=(x,\bx_\perp)^{\ms T}$, where $\bx_\perp$ is an $n$-dimensional vector. Similarly, the momentum is written as $\bp=(p ,\bp_\perp)^{\ms T}$. 

As our $D$ dimensional  Minkowski wavepacket we take a Minkowski plane wave with a Gaussian envelope. The spatial extension of the envelope in the parallel direction is denoted by $\si$. We will refer to this as the \emph{length} of the wavepacket. Additionally, the width in the $n$ perpendicular directions is taken to be equal, and is denoted by $\si_\perp$. We simply call this \emph{the width}. For a depiction of the parameters $\si$ and $\si_\perp$, see figure \ref{fig:MinkowskiWavepacket}(b). 

With these definitions in place, our $D$-dimensional wavepacket mode reads 
\begin{align}
	\label{eq:initialValueConditionsOfDDimensionalMinkowsiWavepacket}
	\begin{aligned}
g_k(0,\bx)&=\frac{1}{\mc N}\exp\left(-\frac{|\bx_\perp|^2}{2\si_\perp^2}-\frac{x ^2}{2\si ^2}+i  k  x \right),\\
(\del_t g_k)(0,\bx)&=\left(-i  k +\frac{x }{\si ^2}\right) g_k(0,\bx).
\end{aligned}
\end{align}
This is a natural extension of the initial value conditions of a 1+1 dimensional Gaussian wavepacket (\ref{eq:MinkowskiWavepacket}). Again, $\mc N$ is some (new) normalization constant. 


Having defined the wavepacket we now move on to the calculation of the entropy. Although we will skip the calculations, an important note is in order. As mentioned before, the expectation values are ultimately some integral of the mode functions. In the present case however, these integrals cannot be solved completely. Instead, we have used a saddle-point-like method and found the relevant expectation values as an asymptotic series. The full series can be found in section \ref{sec:calcsMinkowskiD}.

The first few terms of the relevant expectation values of a $D$-dimensional Gaussian Minkowski wavepacket are
\begin{align}
\label{eq:MinkowskiDExpectationValues}
	\langle \hat b_k^\dagger \hat b^\nodagger_k \rangle &= \frac{n(n+2)}{64\,( k  \si_\perp)^4}\left(1+\frac{3}{( k  \si )^2}+\frac{45}{4( k  \si )^4}\right)+\mathcal O \left[\frac{1}{( k \si )^6( k \si_\perp)^4}\right],\\
\langle \hat b_k \hat b_k \rangle  & = \mathcal O\left[\e^{-( k \si )^2}\right].\nn
\end{align}
The symplectic eigenvalue is given directly by (\ref{eq:symplecticEigenvalueOfOneMode}), and plugging this into (\ref{eq:entropy}) yields the entropy. The resulting expression is a bit unwieldy, but since the entropy $S$ is a monotonically increasing function of $\langle \hat b_k^\dagger \hat b^\nodagger_k \rangle$, we can look at the latter to qualitatively discuss the entropy. 

Let us first check consistency with the 1+1 dimensional result of the previous subsection. We can go from the $D$-dimensional result  we have here to the 1+1 dimensional result in two ways. Firstly, we can put $n$, the number of perpendicular directions, to zero. Secondly, we can send $k\si_\perp\to\infty$ because an infinitely wide wavepacket is in essence a 1+1 dimensional wavepacket. In both cases, $\langle \hat b_k^\dagger \hat b^\nodagger_k \rangle$ goes to zero. This is consistent because the 1+1 dimensional result is exponentially small, and we neglect exponentially small contributions in (\ref{eq:MinkowskiDExpectationValues}).

There is another reason to look at the limit $k\si_\perp\to\infty$. As mentioned before, in this limit the $D$-dimensional wavepacket behaves in essence as a 1+1 dimensional wavepacket. This also means that if we sequentially send $k\si\to 0$, then $S\approx 0.35\,$bit. Thus we conclude that a Gaussian Minkowski wavepacket that is infinitely wide, but has a vanishing length, has an entropy of $S\approx 0.35\,$bit. 

A final thing to observe from (\ref{eq:MinkowskiDExpectationValues}) is the asymmetric behavior between $k\si$ and $k\si_\perp$: when we send $k\si\to\infty$ while keeping $k\si_\perp$ fixed, the entropy assumes some finite value. Contrastingly, when we send $k\si_\perp\to\infty$ and keep $k\si$ fixed, the entropy vanishes. Or, in other words, a spaghetti-like wavepacket has more localization entropy than a pancake-like wavepacket. 

\section{A pair of Rindler wavepackets}
\label{sec:aPairOfRindlerWavepackets}
We now move to Rindler space (see section \ref{sec:RindlerSpace}). As in the previous section, we start in 1+1 dimensions and then generalize to $D$ dimensions. In both cases, we compute the entropy of a single Rindler wavepacket (i.e. $N=1$), and a pair of Rindler wavepackets (i.e. $N=2$). These two quantities will tell us how much the Rindler wavepackets are entangled. More details on the calculations can be found in section \ref{sec:calcRindler}. In this section and section \ref{sec:calcRindler} we put $a=1$.
\subsection{1+1 dimensions}
As our pair of Rindler wavepackets, we take a pair of Rindler plane waves (\ref{eq:RindlerPlaneWaves}), both with a Gaussian envelope in $\xi$,
\begin{align}
	\label{eq:rindlerWavepacket}
	\begin{aligned}
	h^R_{k}(\eta,\xi)&=\frac{1}{\mc N}\exp\left[-\frac{1}{2\si^ 2}(\xi-\eta)^2+ik(\xi-\eta)\right],\\
	h^L_{-k}(\eta,\xi)&=\frac{1}{\mc N}\exp\left[-\frac{1}{2\si^ 2}(\xi-\eta)^2-ik(\xi-\eta)\right]	.
	\end{aligned}
\end{align}
Here $\mc N$ is some (new) normalization constant, $(\eta,\xi)$ the Rindler coordinates in the appropriate wedge, $k>0$ the approximate Rindler Momentum, and $\si$ the length in Rindler coordinates. Both modes are positive-frequency and right-moving. Note that strictly speaking we have again given more information about our wavepacket than necessary since we will only be using the initial value conditions. The mode operators $\hat d^{\ms{R}}_k$ and $\hat d^{\ms L}_{-k}$ are associated with the Rindler wavepacket modes above. A schematic depiction of the $D$-dimensional extension of these modes can be found in figure \ref{fig:pairOfRindlerWavepackets}.

\paragraph{The entropy of one Rindler Wavepacket}
Again, a saddle-point-like method was used to obtain the relevant expectation values. The explicit calculation can be found in section~ \ref{sec:calcRindler}.

We found that the relevant expectation values of a single Rindler wavepacket that is in either $R$ or $L$ equals
\begin{equation}
\label{eq:expectationValuesOneRindlerWavepacket}
\begin{aligned}
\langle \hat d^{\ms R}_k \hat d^{\ms R}_k \rangle=\langle \hat d^{\ms L}_{-k}\hat d^{\ms L}_{-k} \rangle&=\mathcal O \left[\e^{-(k\si)^2}\right],\\
\langle \hat d^{\ms R \dagger}_k \hat d^{\ms R \nodagger}_k\!\!\rangle =\langle \hat d^{\ms L \dagger}_{-k}\hat d^{\ms L\nodagger}_{-k} \rangle   &=N_k\left[1+\frac{\varphi_k}{ \si^2}+\mathcal O\left(\frac{1}{\si^4}\right)\right],
\end{aligned}
\end{equation}
with $\varphi_k=\frac{\pi}{2}  \left(1+\coth \pi k \right) \left(\pi\, \coth\pi k-1/k\right)$ and $N_k$ the thermal spectrum (\ref{eq:thermalSpectrum}). Remember that for the Rindler plane waves, $\langle \hat b^{\ms R}_p \hat b^{\ms R}_p \rangle=\langle \hat b^{\ms L}_{-p}\hat b^{\ms L}_{-p} \rangle=N_p$ (also see eq. \ref{eq:thermalSpectrum}).

\begin{figure}[h]
\centering
\def\svgwidth{1\textwidth}
\smaller{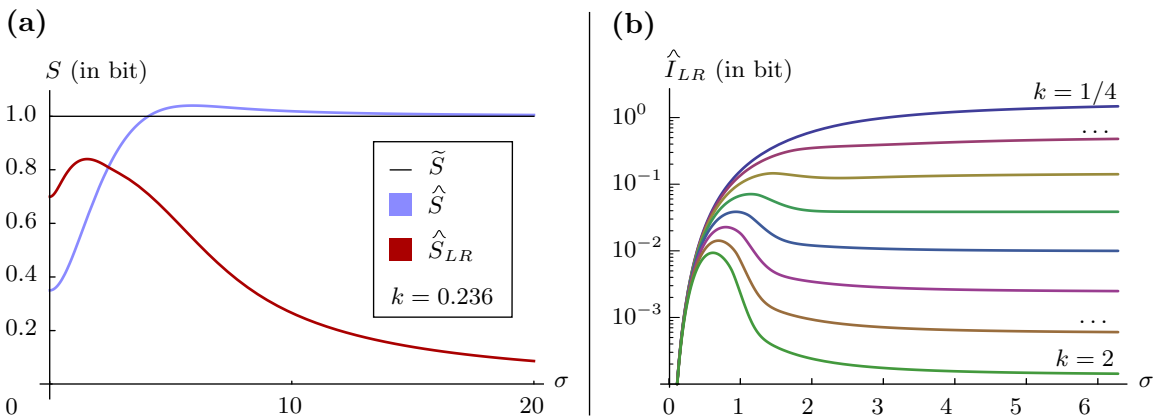}
\caption{
\textbf{(a)} The entropy of: a Rindler wavepacket, a pair of Rindler wavepackets and a Rindler plane wave; all as a function of the length $\si$. The Rindler momentum $k$ is chosen as to make $\widetilde S\approx 1\,$bit. \textbf{(b)} Log-lin plot of the mutual information of two Rindler wavepackets for various $k$ with increments of $1/4$, as a function of the length $\si$. 
\label{fig:numericalResults}}
\end{figure}

By computing the symplectic eigenvalues and inserting these into the entropy, we find the entropy of a single 1+1 dimensional Rindler wavepacket equals
\begin{align}
\label{eq:entropyOfRindlerWavepacket}
\wavepacket S_R=\wavepacket S_L\equiv \wavepacket S=\widetilde S+\frac{2\pi N_k \varphi_k}{\si^2} +\mc O\left(\frac{1}{\si^4}\right),
\end{align}
where $\widetilde S$ is the entropy of an Rindler plane wave (\ref{eq:entropyOfRindlerPlaneWave}). We use the symbol $\!\mathlarger{\wavepacket {}}$ to denote wavepackets. Next to this analytic result, we have computed $\wavepacket S$ numerically. For the results, see figure \ref{fig:numericalResults}(a).

Additionally, we found that in the limit $\si\to 0$, $s=1/\sqrt{\pi}$, as was also the case for the 1+1-dimensional Minkowski wavepacket. So an infinitely narrow Rindler wavepacket also has an entropy of approximately $0.35\,$bit.

\paragraph{The entropy of a pair of Rindler wavepackets}

We now move to the entropy of the combined system of $h_k^{\ms R}$ and $h_{-k }^\ms{L}$, so here $N=2$. Calculating the $4\times 4$ covariance matrix, we find the relevant expectation values are the ones we already found (\ref{eq:expectationValuesOneRindlerWavepacket}), together with
\begin{equation}
\label{eq:expectationValuesOfAPairOfRindlerWavepackets}
\begin{aligned}
\langle \hat d_k^{\ms R} \hat d_{-k}^{\ms L} \rangle  &=\frac{1}{2\sinh\pi k}\left[1+\frac{\vartheta_k}{\si^2}+\mathcal O\left(\frac{1}{\si^4}\right)\right],\\
\langle \hat d_k^{\ms R\dagger}\hat d_{-k}^{\ms L\nodagger}\rangle  &=\mathcal O\left[e^{-(k\si)^2}\right],
\end{aligned}
\end{equation}
where $\vartheta_k=\frac{\pi}{2}\left[\pi  \coth ^2 \pi k -\frac{ \coth \pi  k}{k}-\frac{\pi }{2}\right].$
Computing the symplectic eigenvalues and plugging them into (\ref{eq:entropy}) yields
\begin{align}
\label{eq:entropyOfAPairOfRindlerWavepackets}
	\wavepacket{S}_{LR}= \frac{1+\log( 2\kappa^2\si^2)}{\kappa^2\si^2}+\mathcal O \left[\frac{1}{\si^4}\right],
\end{align}
with $\kappa=\frac{2}{\pi} \sinh \pi k$.
Again, a plot of the numerical result can be found in figure \ref{fig:numericalResults}(a).

\paragraph{The Mutual information of two Rindler wavepackets}
The mutual information between subsystems $A$ and $B$ is defined as $I_{AB}=S_A+S_B-S_{AB}$, where $S_{X}$ is the von Neumann entropy of system $X$. The entropy of one Rindler plane wave equals $\widetilde S=\widetilde S_R=\widetilde S_L$ as in (\ref{eq:entropyOfRindlerPlaneWave}).   Furthermore, $\widetilde S_{LR}=0$ (see eq. \ref{eq:entropyOfAPairOfRindlerPlaneWaves}). Therefore, the mutual information of two Rindler plane waves equals $\widetilde I_{LR}=2 \widetilde S$.

The localization of the Rindler plane waves introduces corrections to this result. With the results from the previous two paragraphs, the mutual information of two 1+1 dimensional Rindler wavepackets equals
\begin{align}\label{eq:mutualInformationOfTwoRindlerWavepackets}
	\wavepacket I_{LR}&=\widetilde I+\frac{1}{\si^2}\left(4\pi N_k\varphi_k-\frac{1+\log(2\ka^2\si^2)}{\ka^2}\right)+\mathcal O \left(\frac{1}{\si^4}\right),
\end{align}
with $N_k$, $\varphi_k$ and $\kappa$ functions of $k$, as defined before. 
In addition, numerical results can be found in figure \ref{fig:numericalResults}(b). 

A feature that stands out in the numerical results, is that $\wavepacket I_{LR}$ can not only be smaller than the asymptotic value, but also larger. For small $k$ it can actually be much larger, with a peak at around $\si=1$.

This behavior is explained by our analytical result (\ref{eq:mutualInformationOfTwoRindlerWavepackets}). There are two corrections to the asymptotic result with competing effects, represented by the two terms in the parenthesis. On the one hand, we have the first term, $4\pi N_k\varphi_k$. It originates from the one-sided entropies $\wavepacket S_R$ and $\wavepacket S_L$, and increases the mutual information. As is clear form section \ref{sec:MinkowskiWavepacket}, localization alone can introduce some entropy. So, when two Rindler plane waves are localized, there is simply more entropy to be shared, which can lead to an increase of the mutual information. 

On the other hand, we have the second term. It originates from the two-sided entropy $\wavepacket S_{LR}$, and decreases the mutual information. As we have seen in (\ref{eq:entropyOfAPairOfRindlerPlaneWaves}), the Rindler plane waves purify each other perfectly. These plane waves are a very special choice. If we localize the plane waves, they become less like the Rindler plane waves, so we can expect their mutual information to decrease. 

\subsection{$D$ dimensions}\label{sec:aPairOfDDimensionalRindlerWavepackets} In this subsection we extend the results of the previous to $D$ dimensions, with $D> 1+1$. Analogous to the extension of the 1+1 dimensional Minkowski wavepacket, we define our $D$-dimensional Rindler wavepackets by 
\begin{equation}
\label{eq:DDimensionalRindlerWavepacket}
\begin{aligned}
I^{(\gamma)}_k(0,\bx)&=\frac{1}{\mc N}\exp\left[{-\frac{\bx^2_\perp}{2\sigma^2_\perp}-\frac{\xi^2}{2\si ^2}+i k\xi}	\right],\\
	\del_{\eta}  I^{(\gamma)}_k(0,\bx)&=\left(-i\gamma k+\frac{\xi}{\si ^2}\right) I^{(\gamma)}_k(0,\bx),   
\end{aligned}
\end{equation}
with $\mc N$ some (new) normalization,  $k>0$, $\gamma=1$ in $R$, $\gamma=-1$ in $L$ and $\bx=(\xi,\bx_\perp)^T$. 

Again, an infinitely wide, $D$-dimensional wavepacket is in essence a 1+1 dimensional wavepacket. In other words, in the limit $\si_\perp\to\infty$, the relevant expectation values of a (pair of) $D$-dimensional Rindler wavepacket(s) are given by (\ref{eq:expectationValuesOneRindlerWavepacket}, \ref{eq:expectationValuesOfAPairOfRindlerWavepackets}), the entropy by (\ref{eq:entropyOfRindlerWavepacket}, \ref{eq:entropyOfAPairOfRindlerWavepackets}) and the mutual information by (\ref{eq:mutualInformationOfTwoRindlerWavepackets}). Accordingly, a $D$-dimensional Rindler wavepacket has an entropy of approximately $0.35\,$bit in the limit $\si_\perp\to\infty$, $\si\to 0$. There is no dependence on $D$ in any of the aforementioned quantities, as is more clear in section \ref{sec:calcRindler}.

Decreasing $\si_\perp$ from infinity introduces corrections to the 1+1-dimensional results. Since $\si$ is measured in Rindler coordinates, naively one would expect these corrections to be small whenever $e^\si/\si_\perp\ll 1$. However, as a lengthy and technical calculation shows (section \ref{sec:calcRindlerDdimensions}), the relevant quantity that controls the corrections to the expectation values is 
\begin{equation*}
\om\equiv\frac{\log 2\si_\perp}{\si^2},
\end{equation*}
and corrections are always or order $e^{-c\om}$, with $c>1$ if $\om>1$.

As an example, let us look at the corrections to the expectation value of the number operator,
\begin{align*}
\ddd=N_k\left[1+\frac{\varphi_k}{\si^2}+\mathcal O\left(\frac{1}{\si^4}\right)\right]+
e^{\si^2(1-k^2)}\times
\left\{\begin{array}{ll}
\mathcal O [\si e^{\si^2(1-2\om)}]\\
\mathcal O [ e^{-\si^2\om^2}]\\
\mathcal O [\si e^{-\si^2 2\om}]
\end{array}\right. ,
\end{align*}
with $\varphi_k$ as before (cf. eq. \ref{eq:expectationValuesOneRindlerWavepacket}). In the piecewise definition of the correction, we have assumed $k<1$. For this expectation value, we can see there are two regimes that are separated by $\om_{crit}\equiv 1-k^2/2$. Corrections are exponentially large for $\om>\om_{crit}$, but exponentially small for $\om<\om_{crit}$. 

From the above we can conclude that one cannot assume a Rindler wavepacket to be a Rindler plane wave whenever $\om\lesssim 1$. The exponential difference in the expectation values of the respective number operators shows they are really different modes. 

Note there is an asymmetry between the two directions, as is also the case for the Minkowski wavepacket. If we first take $\si_\perp\to\infty$, the wavepacket is like a 1+1 dimensional wavepacket for any $\si$. If we, however, first send $\si\to\infty$, the expectation value of the number operator is infinite for any finite $\si_\perp$.  

So there are two surprising facts. First of all $\si_\perp$ has to be tremendously large for the corrections to be negligible. Secondly, the transition to the regime where the corrections are no longer negligible is very sudden, and it behaves as a phase transition. It would be interesting to study this further as a phase transition, possibly in the context of RG-flows. 

\section{Application to the firewall paradox}
\label{sec:applicationToTheFirewallParadox}
With the exception of the introduction, there has been no reference to the Firewall paradox. This is because the results are interesting in their own right, and to be clear that they do not depend on the details of the paradox in any way. In this section we comment on how the results can be used in the discussion of the firewall paradox. 

The firewall paradox \cite{AMPS}, seems to show a contradiction between unitarity, equivalence and effective field theory. There are many formulations of the paradox, and we will give our own here to be precise on how our results fit in, although a comprehensive formulation is out of the scope of this article.  Our formulation is mostly adapted from \cite{HarlowHayden}, and more details can be found there. 

\subsection{The paradox}
Consider a black hole that was formed by a collapsing shell of photons that are together in a pure state. Say we have waited long enough so that the majority of degrees of freedom have now escaped the black hole due to Hawking radiation. 

\begin{figure}[h]
	\centering
	\def\svgwidth{1\textwidth}
	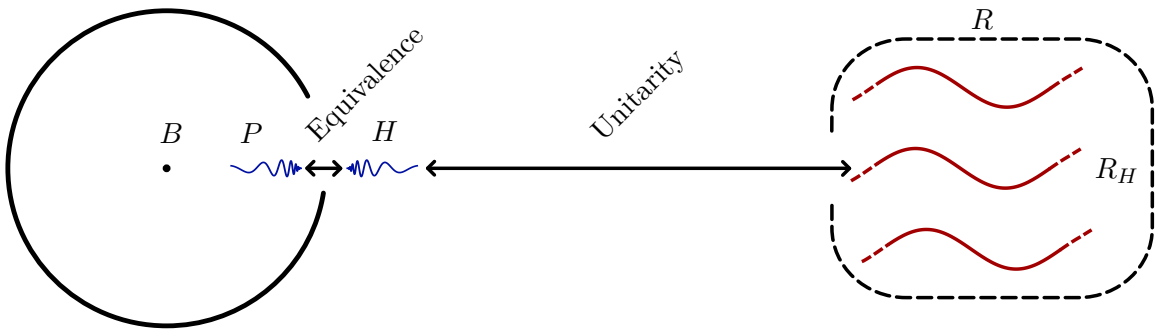\caption{Schematic depiction of the various subsystems at play in the firewall paradox. Arrows symbolize entanglement. Definitions of the other symbols are given in the text. \label{fig:firewallParadox}}
\end{figure}

Now let $H$ be a newly emitted, outgoing Hawking mode that is still close to the horizon, $P$ its partner mode behind the horizon, $B$ the remaining degrees of freedom of the black hole, and $R$ the far-away radiation that has long escaped the black hole. For a schematic depiction of the various subsystems, see figure \ref{fig:firewallParadox}. 

Firstly, by the equivalence principle, space should locally look like Minkowski space, in particular near a black hole horizon. Near-horizon Hawking modes are like Rindler modes \cite{AMPS, Harlow, Susskind}. These Rindler modes need to be entangled with their partners, for otherwise space does locally not look like the Minkowski vacuum (see eq. \ref{eq:minkowskiVacuumInTermsOfRindlerModes}). As we have seen in section \ref{sec:RindlerSpace}, a Rindler mode by itself has some order one entropy, whereas a pair of Rindler modes has no entropy. In other words, 
\begin{equation*}
	S_H-S_{PH}\sim 1.
\end{equation*}

Secondly, because we started out in a pure state, we have by unitarity that the whole ($B$, $P$, $H$ and $R$ together) is in a pure state. Since the abundance of degrees of freedom is in $R$, and the black hole formation and evaporation process is so complex that is scrambles all information evenly over the degrees of freedom, there is a sub-factor $R_H$ in $R$ that purifies $H$ to a large extent \cite{Page}. In other words, for the entropy of the combined system $HR_H$ we have
\begin{equation}
S_{HR_H}\ll 1.
\end{equation}
Since $H$ alone is in a mixed state this means $H$ and $R_H$ are entangled.

Now comes the paradox. Any three quantum systems obey strong subadditivity of their von Neumann entropy \cite{Lieb},
 \begin{align}
 	\label{eq:strongSubaditivity}
 	S_{H{R_H}}-S_{PH{R_H}}\geq S_H-S_{PH}.
 \end{align}
 As we have argued, the right hand side is of order unity, whereas $S_{HR_{H}}$ is much smaller than unity, and $S_{PHR_{H}}$ is positive. This is clearly a contradiction. 

We can rewrite the above to get a statement that involves the mutual information. By (normal) subadditivity and the triangle inequality, 
\begin{align*}
	|S_P-S_{HR_H}|\leq S_{PHR_H}\leq S_P+S_{HR_H},
\end{align*}
so that  $S_{PHR_H}=S_P+\ep$ with $|\ep|\ll 1$. Substituting this into (\ref{eq:strongSubaditivity}), we get
\begin{align}
\label{eq:unitarity}
	S_{HR_H}-\ep\geq I_{HP},
\end{align}
with $I_{HP}$ the mutual information between $H$ and $P$. We will explicitly show that our Rindler wavepackets have a mutual information that violates this inequality.  

\subsection{Rindler-Hawking wavepackets}\label{sec:RindlerHawking}
In the formulation above, but certainly also in \cite{AMPS}, it is argued that close to the horizon, space is like Rindler space, and near-horizon Hawking modes are like Rindler modes. However, Rindler modes are infinitely extended, but by definition,  near-horizon Hawking modes are not. Moreover, even though space might locally look like Rindler space, the Rindler modes extend into the region where the Schwarzschild metric is no longer approximated by the Rindler metric. So in a more rigorous treatment, one would have to consider localized Rindler modes.

However, the localization induces some entropy, and this entropy need not be shared with the mirror mode on the other side of the horizon. So a priori, it is not clear that a localized Rindler mode is purified by its partner enough. Luckily, with the results of the previous sections at hand, we can show that under certain conditions, this is indeed possible. 

To do so, let $2GM$ be the Schwarzschild radius of an old black hole, and say we trust the Rindler approximation up to a proper distance $\rho_{\rm{max}}$ from the horizon. As our near-horizon Hawking wavepacket $H$ we can now take a Rindler wavepacket, as defined in this paper, that runs from $\rho=\frac{1}{a}e^{-na\si}$ to $\rho=\frac{1}{a}e^{na\si}$ with $\frac{1}{a}e^{na\si}=\rho_{\rm_{max}}$. Here $n$ determines how strict a confinement we demand. Remember that if $\si_\perp>\frac{1}{2a}e^{(a\si)^2}$, a $D$-dimensional wavepacket is in essence 1+1 dimensional (section \ref{sec:aPairOfDDimensionalRindlerWavepackets}). For any $n$, $\si$ and $\rho_{\rm{max}}$ (and in particular for large $\si$ so that a 1+1 dimensional Rindler wavepacket is purified by its partner), there exists a $2GM$ and $\si_\perp$ so that $2GM\gg \si_\perp > \frac{1}{2a} e^{(a\si)^2}$ while respecting $\frac{1}{a}e^{-n\si}\gg l_p$ and $\frac{1}{a}e^{n\si}=\rho_{\rm{max}}$. Here $l_p$ is the Planck length. 

In other words, there are certain conditions to be met for the pair of Rindler wavepackets to be highly entangled, and the size of the black hole can always be increased so that these conditions are met with arbitrary satisfaction. 

As we have shown in this paper (in particular, see eq. \ref{eq:mutualInformationOfTwoRindlerWavepackets} and fig. \ref{fig:numericalResults}(b)), a pair of Rindler wavepackets has $\wavepacket I_{LR}\sim 1$, while on the other hand, unitarity demands $\wavepacket I_{LR}\sim 0$ (eq. \ref{eq:unitarity}). This, exactly, is the paradox. 

\section{Conclusion}
\label{sec:conclusion}
In this paper, a recipe was given to compute the entanglement entropy of any mutually orthogonal set of modes in the Minkowski vacuum. This recipe was applied to a Minkowski plane wave with a Gaussian envelope, and a pair of Rindler plane waves, both with a Gaussian envelope. These results enabled us to gain more insight in the entanglement structure of the Minkowski vacuum, and explicitly construct the Hawking wavepackets that play an essential role in the firewall paradox. 

For a $D$-dimensional Minkowski wavepacket, we found the entropy goes to zero as a power law in the spatial extension of the wavepacket. There is an asymmetry in this power law: a spaghetti-like wavepacket has more entropy than a pancake-like wavepacket. In addition, we found the entropy obtains a specific parameter-independent value of approximately $0.35\,$bit in the limit of infinite width but vanishing length (i.e. $\si_\perp\to\infty$, $\si\to 0$). 

For the $D$-dimensional Rindler wavepackets, we found that they are essentially 1+1 dimensional if $\om\equiv \log(2\si_\perp)/\si^2>1$. For a $1+1$ dimensional wavepacket, the corrections to the entropies are a power law (for the one-sided entropy) or slightly worse than a power law (for the two-sided entropy). A $D$-dimensional Rindler wavepacket also obtains an entropy of approximately $0.35\,$bit in the limit of infinite width and vanishing length. 

If, however, $\om<\om_{crit}\equiv 1-k^2/2$, the $D$-dimensional Rindler wavepacket is nothing like a $1+1$ dimensional wavepacket. This is shown by the expectation value of the number operator, which goes like $e^{-c \om}$ if $\om>\om_{crit}$, but like $e^{c \om}$ if $\om<\om_{crit}$. (In both cases $c$ is some positive constant.) 

Some questions still remain. 
First of all, we have treated gravity as a fixed background. It would be interesting to see whether gravitational back-reaction could be included, and whether it could be of any influence on our conclusions. 

Also, there are two things we as of yet do not have any physical intuition for. Firstly, we found an entropy of approximately $0.35\,$bit in the limit of infinite width and vanishing length for both the Minkowski wavepacket and the Rindler wavepacket. This suggests that this value might be obtained by any highly localized mode in the same limit. It would be interesting to find out what physical principle is behind this parameter-independent, dimensionless constant. 

Secondly, we have no intuition for why the ratio $\om=\log(2\si_\perp)/\si^2$ turns out to the natural parameter for describing the corrections to the expectation values of the 1+1 dimensional Rindler wavepacket. Also, the behavior of the expectation value of the number operator hints there is a phase transition at $\om_{crit}=1-k^2/2$. It would be interesting to see if this behavior can be properly described as a phase transition, and if so, we could ask: why at $1-k^2/2$?

\clearpage

\part{Calculations}
This part is structured as sections  \ref{sec:MinkowskiWavepacket} and \ref{sec:aPairOfRindlerWavepackets} and shows the calculations. Although technical and lengthy, the calculation in section \ref{sec:calcExValDDimensionalRindlerWavepakets} gains some insight in the phase transition-like behavior of the expectation values of a $D$-dimensional Rindler wavepacket because mathematically we can pinpoint where and why the transition occurs.  

\section{Minkowski wavepacket}
\label{sec:calcMinkowski}
\subsection{1+1 dimensions}
The initial value conditions of a Minkowski plane wave with momentum $p$ are
\begin{equation*}
\begin{aligned}
f_p(0,x)&=\frac{1}{\sqrt{4\pi\om_p}}\e^{i p x},\\
(\partial_t  f_p)(0,x)&=(-i\om_p) f_p(0,x),
\end{aligned}
\end{equation*}
with $\om_p=|p|$. From the full definition of the Rindler wavepacket (\ref{eq:MinkowskiWavepacket}), we have after normalization,
\begin{equation*}
\begin{aligned}
g_k(0,x)&=\frac{1}{\pi^{1/4}\sqrt{2k\si}}\exp\left(-\frac{x^2}{2\si^2}+i k  x\right),\\
(\del_t  g_k)(0,x)&=\left(-i k +\frac{x}{\si^2}\right)g_k(0,x).
\end{aligned}
\end{equation*}
The Bogolyubov coefficient $\alpha_{kp}$ that is needed to write the wavepacket in terms of the plane waves is then found by (see eq. \ref{eq:alphaBeta})
\begin{align}
\label{eq:alphaMinkowskiWavepacket}
	\al_{kp}&=(g_k,f_p)\\
&=-i \integral x\, g_k(0,x) f^*_p(0,x)\left(i\om_p+i k -\frac{x}{\si^2}\right).
\end{align} 
By solving this integral we find
\begin{align}
	\label{eq:betaMinkowskiWavepacket}
	\al_{kp}&=\left\{\begin{array}{ll}0&:p\leq0\\\sqrt{\frac{\si p}{ k }}\pi^{-1/4}\exp[-\frac{\si^2}{2}(k-p)^2]&:p>0\end{array}\right..
	\intertext{Likewise, we obtain for $\be_{kp}=-(g,f^*)$,}\label{eq:wp1beta}
	\be_{kp}&=\left\{\begin{array}{ll}0&:p\leq0\\\sqrt{\frac{\si p}{ k }}\pi^{-1/4}\exp[-\frac{\si^2}{2}(k+p)^2]&:p>0\end{array}\right..
\end{align}
The resulting expectation values are given in the text.

\subsection{$D$ dimensions}
\label{sec:calcsMinkowskiD}
To save writing, we will work with unnormalized functions. As a reminder, we will put tildes over these functions. These tildes should not be confused with the tildes in part I, which denote plane waves. 

In $D$-dimensions, the initial value conditions of an unnormalized Minkowski plane wave are 
\begin{align*}
	\tilde f_\bp(0,\bx)=\e^{i \bp\cdot\bx},&&(\partial_t \tilde f_\bp)(0,\bx)=(-i\om_\bp)\tilde f_\bp(0,\bx),
\end{align*}
with $\om_\bp=\sqrt{p ^2+|\bp_\perp|^2}$. The initial value conditions for the wavepacket are (\ref{eq:initialValueConditionsOfDDimensionalMinkowsiWavepacket}).

\paragraph{Normalization}
To properly normalize the wavepacket, we compute the norm of the unnormalized wavepacket 
\begin{align*}
\mc N^2=(\tilde g_k,\tilde g_k)=2\pi^{d/2} k \si \si_\perp^n.
\end{align*}
The normalized mode function is then given by $g_k=\tilde g_k/
	\mc N$.

The Minkowski plane waves are delta function normalized, 
\begin{align*}
	(\tilde f_\bp,\tilde f_{\bp'})_\delta=2\om_\bp(2\pi)^{d}\delta(\bp-\bp'). 
\end{align*}
Let us define $(\tilde f_\bp,\tilde f_{\bp'})\equiv(\tilde f_\bp,\tilde f_{\bp'})_\delta/\delta(\bp-\bp')$ so that we can write the normalized plane waves as $f_\bp=\tilde f_\bp/\sqrt{(\tilde f_\bp,\tilde f_\bp)}$.

\paragraph{Bogolyubov coefficients}
By (\ref{eq:alphaBeta}),
\begin{align*}
	\al_{k\bp}&=\frac{1}{2}\pi^{-\frac{1}{4}(n+1)}\sqrt{\frac{\si \si_\perp^n}{ k \,\om_\bp}}\,(p +\om_\bp)\exp\!\left[-\frac{\si_\perp^2}{2}|\bp_\perp|^2-\frac{\si ^2}{2}(p - k )^2\right],\\
	\be_{k\bp}&=\frac{1}{2}\pi^{-\frac{1}{4}(n+1)}\sqrt{\frac{\si \si_\perp^n}{ k \,\om_\bp}}\,(p +\om_\bp)\exp\!\left[-\frac{\si_\perp^2}{2}|\bp_\perp|^2-\frac{\si ^2}{2}(p + k )^2\right].
\end{align*}
Note that the only difference between $\al_{k\bp}$ and $\be_{k\bp}$ is the sign in front of the $ k $ that is in the exponent.

\paragraph{Expectation values}
The expectation values that go into the symplectic eigenvalue are $\langle \hat b_k^\dagger \hat b^\nodagger_k\rangle $ and $\langle \hat b_k \hat b_k\rangle $ (see eq. \ref{eq:symplecticEigenvalueOfOneMode}). The former equals
\begin{align*}
	\langle \hat b_k^\dagger \hat b^\nodagger_k\rangle &=\integral\bp| \beta_{k\bp}|^2\nn\\
    &=\frac{\si \,\si_\perp^{n}}{4\,\sqrt{\pi^{n+1}}}\,S(n-1)\int_{\mathsmaller{-\infty}}^\mathsmaller{\infty} \!\!\!\!\mathrm{d} p \!\!\int_{\mathsmaller{0}}^{\mathsmaller{\infty}}\!\!\!\! \mathrm d|\bp_\perp|\, |\bp_\perp|^{n-1}\, \frac{(p +\om_\bp)^2}{ k \,\om_\bp}{\mathlarger{e}}^{-\si_\perp^2|\bp_\perp|^2-\si ^2(p + k )^2}.
\end{align*}
Here, $S(n-1)$ is the surface area of a unit $n-1$ sphere.

Unfortunately, we cannot solve the last integral due to the factor $\frac{(p +\om_\bp)^2}{ k \,\om_\bp}$. To tackle this, we expand this factor around the part where the rest of the integrand is exponentially peaked. The integral that is obtained by plugging in this expansion \emph{can} be solved. Note that the width of the exponential peak scales as $1/\si_\perp$ in one direction, and as $1/\si$ in the other, so that the approximation becomes better with increasing $\si$ and $\si_\perp$. By the above procedure, we will obtain a solution to the integral as an asymptotic series. 

Computing many terms of the expansion, we find they coincide with
\begin{align*}
	\frac{(p +\om_\bp)^2}{ k \,\om_\bp}\approx\sum_{\ell=0}^{L}\sum_{j=0}^J C_{\ell j}\left(\frac{|\bp_\perp|}{ k }\right)^{2\ell+4}\left(\frac{p + k }{ k }\right)^j,
\end{align*}
where
\begin{align*}
	C_{\ell j}&=\frac{1}{\sqrt{\pi}}\frac{(-1)^\ell}{(2+\ell)!\,\ell !}\frac{2^j}{j!}\Gamma\left(a_j^\mathsmaller{(-)}+\ell+1\right)\Gamma\left(a_j^\mathsmaller{(+)}+\ell+1\right),\\
	a_j^\mathsmaller{(\pm)}&=\frac{1}{4}(3\pm(-1)^j+2j).
\end{align*}
So we find that 

\begin{align*}
\bdb_{LJ}&=\frac{n}{8\sqrt{\pi}}\sum_{\ell=0}^L\sum_{j=0}^J\frac{(-1)^\ell}{j!}(1+\ell)_{1+j}(1+\sfrac{n}{2})_{1+\ell}(3+\ell)_{j-\frac{3}{2}}\ ( k  \si_\perp)^{-4-2\ell}( k \si )^{-2j},
\end{align*}
where we used the Pochhammer symbol $(a)_n=\Gamma(a+n)/\Gamma(a)$ as shorthand notation. To make this result a bit more insightful, we note that he first few terms read
\begin{align*}
\bdb=\frac{n(n+2)}{64\,( k  \si_\perp)^4}\left(1+\frac{3}{( k  \si )^2}+\frac{45}{4( k  \si )^4}\right) +\mc O \left(\frac{1}{\si^6 \si_\perp^4}\right).
\end{align*}

The second relevant expectation value is 
\begin{align*}
	\langle \hat b_k \hat b_k \rangle =-\integral \bp\,\al_{k\bp}^* \be_{k\bp}^*
	=\mathcal O \left[\mathlarger e^{-( k  \si )^2}\right].
\end{align*}

\section{A pair of Rindler wavepackets}
\label{sec:calcRindler}
\subsection{1+1 dimensions}
The mode function of a 1+1 dimensional Rindler wavepacket is given in equation \ref{eq:rindlerWavepacket}. We associate the operators $\hat d^{\ms R}_k$ and $\hat d^{\ms L}_k$ with the two Rindler wavepackets.

\paragraph{Bogolyubov coefficients}
Here we will take a slightly different approach than before. Instead of writing the Rindler wavepackets in terms of the Minkowski plane waves, we write the Rindler wavepackets in terms of the Rindler plane waves. Sequentially, we use (\ref{eq:rindlerPlaneWavesInTermsOfModesThatAnnihilateTheVacuum }) to compute expectation values. So in this section, the Bogolyubov coefficients are between the Rindler wavepacket and the Rindler plane waves.  

The Rindler metric (\ref{eq:RindlerMetirc}) is conformally equivalent to the Minkowski metric. Therefore, the Klein-Gordon inner product has the same functional form as in Minkowski space (\ref{eq:KleinGordonInnerProduct}). Note also that our wavepacket in the right Rindler wedge (\ref{eq:rindlerWavepacket}) can be obtained from the Minkowski wavepacket (\ref{eq:MinkowskiWavepacket}) by sending $(t,x)\to(\eta,\xi)$. The same holds for the Rindler plane waves.  So, in this section, we can just use the Bogolyubov coefficients (\ref{eq:alphaMinkowskiWavepacket}) and (\ref{eq:betaMinkowskiWavepacket}), but now they act between different modes. 

There are some subtleties concerning the coefficients in the left Rindler wedge $L$. A careful consideration shows 
\begin{align*}
	\al^{\ms R}_{kp}=\al^{\ms L}_{-k-p}\equiv \al_{kp},&& \be^R_{kp}=\be^L_{-k-p}\equiv \be_{kp}.
\end{align*}
So we drop the superscript whenever it is irrelevant. We will also do this for mode operators and expectation values. 

\paragraph{Expectation values} 
Calculating the covariance matrix (\ref{eq:covarianceMatrix}), we find that for our specific choice of modes, it is in the particularly nice form
\begin{align}
	\gv\si=
	\begin{pmatrix}
		A_{00} & 0 & C_{00} & 0\\
		0 & A_{11} & 0 & C_{11}\\
		C_{00} & 0 & A_{00} & 0\\
		0 & C_{11} & 0 & A_{11}\\
	\end{pmatrix},&&\text{with}&&
	\left\{
	\begin{matrix}
		A_{00}&=\ddd+\langle \hat d_k\hat d_k \rangle+\half\\
		A_{11}&=\ddd-\langle \hat d_k\hat d_k \rangle+\half\\
		C_{00}&=\langle \hat d_k^{\ms R\dagger} \hat d_{-k}^L \rangle+\langle \hat d_k^{\ms R} \hat d_{-k}^{\ms L} \rangle\\
		C_{11}&=\langle \hat d_k^{\ms R\dagger} \hat d_{-k}^{\ms L} \rangle-\langle \hat d_k^\ms{R} \hat d_{-k}^{\ms L} \rangle
	\end{matrix}\right. .\label{eq:sigma}
\end{align}
Using equation \ref{eq:bInTermsOfa} (where $\hat a$ and $\hat b$ are now $\hat b$ and $\hat d$ respectively) and (\ref{eq:rindlerPlaneWavesInTermsOfModesThatAnnihilateTheVacuum }), we obtain
\begin{align}
\label{eq:generalExpectationValuesRinderWavepacket}
\begin{aligned}
\langle \hat d_k \hat d_k \rangle
&=-\integral p\ \frac{1}{\sinh(\pi|p| )}\, \al_{kp}^*\be_{kp}^*,
 \\
\langle \hat d_k^\dagger \hat d_k \rangle &=\integral p\,\frac{1}{2\sinh(\pi|p| )}\left(|\al_{kp}|^2e^{-\pi|p| }+|\be_{kp}|^2e^{\pi|p| }\right),
\\
\langle \hat d_k^{\ms R} \hat d_{-k}^{\ms L} \rangle  &=\integral p\,\frac{1}{2\sinh(\pi|p| )}\left(\al^{\ms R*}_{kp}\al^{\ms L*}_{-k-p}+\be^{\ms R*}_{kp}\be^{\ms L*}_{-k-p}\right),
\\
\langle \hat d_k^{\ms R\dagger}\hat d_{-k}^{\ms L}\rangle  &=-\integral p\,\frac{1}{2\sinh(\pi|p| )}\left(\al_{kp}^{\ms R}\be_{-k-p}^{\ms L*}+\be_{kp}^{\ms R}\al_{-k-p}^{\ms L*}\right).
\end{aligned}
\end{align}
We can now insert the Bogolyubov coefficients, which yields
\begin{align}
\label{eq:specificE}
\begin{aligned}
\langle \hat d_k \hat d_k \rangle  &=-\frac{1}{\sqrt{\pi}}\frac{\si}{k}e^{-(k\si)^2}\int_0^\infty\! \!\mathrm{d}p\, \frac{p}{\tanh(\pi\,p )}\, e^{-(p\si)^2},
\\
\langle \hat d_k^\dagger \hat d_k \rangle &=\frac{1}{2\sqrt{\pi}}\frac{\si}{k}\int_0^\infty\!\!\mathrm{d}p\,\frac{p}{\sinh(\pi\, p )}\left(e^{-\si^2(k-p)^2-\pi\, p }+e^{-\si^2(k+p)^2+\pi\, p }\right),
\\
\langle \hat d_k^{\ms R} \hat d_{-k}^{\ms L} \rangle &=\frac{1}{2\sqrt{\pi}}\frac{\si}{k}\int_0^\infty\!\!\mathrm{d}p\, \frac{p}{\sinh(\pi\,p )} \left(e^{-\si^2(k-p)^2}+e^{-\si^2(k+p)^2}\right),
\\
\langle \hat d_k^{\ms R\dagger}\hat d_{-k}^{\ms L}\rangle  &=-\frac{1}{\sqrt{\pi}}\frac{\si}{k} e^{-\left(k\si\right)^2}\int_0^\infty\! \!\mathrm{d}p\, \frac{p}{\sinh(\pi\,p )}\,e^{-\left(p\si\right)^2}.
\end{aligned}
\end{align}

Some of these integrals cannot be solved directly. Like in the previous section, we tackle this by expanding the troublesome parts around the point where the rest of the integrand is exponentially peaked. Note that, again, the width of this peak scales as $1/\si$. Thus we obtain
\begin{equation}
\label{eq:calcExpectationValuesOfAPairOfRindlerWavepackets}
\begin{aligned}
\langle \hat d_k \hat d_k \rangle 
&=\mathcal O\left[e^{-(k\si)^2}\right],
\\
\langle \hat d_k^\dagger \hat d_k \rangle &=\frac{1}{e^{2\pi k }-1}\left[1+\frac{\varphi(k )}{ \si^2}+\mathcal O\left(\frac{1}{\si^4}\right)\right],
\\
\langle \hat d_k^{\ms R} \hat d_{-k}^{\ms L} \rangle &=\frac{1}{2\sinh(\pi\,k )}\left[1+\frac{\vartheta(k )}{\si^2}+\mathcal O\left(\frac{1}{\si^4}\right)\right],
\\
\langle \hat d_k^{\ms R\dagger}\hat d_{-k}^{\ms L}\rangle  &=\mathcal O\left[e^{-(k\si)^2}\right],
\end{aligned}
\end{equation}
with
\begin{equation*}
\begin{aligned}
\varphi_k&=\frac{\pi}{2}  \left(1+\coth \pi k \right) \left(\pi\, \coth\pi k-1/k\right),
\\
\vartheta_k&=\frac{\pi}{2}\left[\pi  \coth ^2 \pi k -\frac{ \coth \pi  k}{k}-\frac{\pi }{2}\right].
\end{aligned}
\end{equation*}
In the limit $\si\to\infty$ the above expectation values are equal to the corresponding expectation values of the Rindler plane waves, as they should. (The latter expectation values can not be found elsewhere in this paper.)

\paragraph{The symplectic eigenvalues}
Since any covariance matrix is real and symmetric, it can be written as \cite{Ferraro}
\begin{align*}
\gv\sigma=
\begin{pmatrix}
\mb A & \mb C\\
\mb C^T & \mb B\\
\end{pmatrix}.
\end{align*}
For a two-mode system, $\mb A$, $\mb B$ and $\mb C$ are real $2\times 2$ matrices. These matrices contain the information about systems $A$, $B$, and the correlations between $A$ and $B$ respectively.

The two symplectic eigenvalues $s_\pm$ are
\begin{align*}
2s^2_\pm=I_0\pm\sqrt{(I_0)^2-4I_4},
\end{align*}
where 
\begin{align*}
I_{0}=I_1+I_2+2I_3,&&
I_1=\det(\mb A),&&I_2=\det(\mb B),&&I_3=\det(\mb C),&&I_4=\det(\gv\sigma).
\end{align*}
Specifically, the symplectic eigenvalues of the covariance matrix (\ref{eq:sigma}), where $\mb B$ happens to be equal to $\mb A$, are thus 
\begin{equation}
s_\pm^2=\left(\!\ddd+\frac{1}{2}\right)^2-\langle \hat d_k^{\ms R} \hat d_{-k}^{\ms L}\rangle^2+\mc O\left(e^{-\si^2k^2}\right).
\end{equation}
Plugging this into ($\ref{eq:entropy}$) and expanding around $1/\si$ gives
\begin{equation*}
\wavepacket S_{LR}=\frac{1+\log(2x^2)}{x^2}+\mc O\left(\frac{1}{\si^4}\right),
\end{equation*}
with $x\equiv 2 \si \sinh(\pi k )/\pi$.

\subsection{$D$ dimensions}
\label{sec:calcRindlerDdimensions}
The initial value conditions of the unnormalized, $D$-dimensional Rindler plane waves are obtained from the full expressions (for example, see \cite{Takagi}),
\begin{equation}\label{eq:initialValueConditionsOfDDimensionalRindlerPlaneWaves}
\begin{aligned}
\tilde	f_\bp^{(\gamma)}(0,\bx)&=\Theta(\rho)K_{i\Om}(|\bp_\perp|\rho)\exp\left[i\,\bp_\perp\!\! \cdot \bx_\perp\right],\\
\del_{\eta}\tilde f_\bp^{(\gamma)}(0,\bx)&=\Theta(\rho)(-i\gamma \Om) 	\tilde f_\bp^{(\gamma)}(0,\bx),
\end{aligned}
\end{equation}
where $\bp=(\Om,\bp_\perp)^T$ with $\Om> 0$,  $\eta$ the Rindler time, $\Theta$ the Heaviside step function, $K_{i\Om}$ the modified Bessel function of the second kind and $\rho=e^{\xi}$ is the proper distance to the horizon. The expression for the left (right) wedge is obtained by setting $\gamma=-1$ $(\gamma=1)$. The initial value conditions of the Rindler wavepacket are given in the text (eq. \ref{eq:DDimensionalRindlerWavepacket}).

\paragraph{Normalization}
The norm of the unnormalized wavepacket equals
\begin{align*}
\mc N^2 = (\tilde I_k,\tilde I_k)&=2 k  (\sqrt{\pi}\si )(\sqrt{\pi}\si_\perp)^n,
\end{align*}
with $k>0$ and $n$ the number of perpendicular directions.
The norm of the Rindler plane waves is 
\begin{align*}
(\tilde f_\bp,\tilde f_{\bp'})_\delta&=\pi^2(2\pi)^n \csch(\pi\Om) \delta(\bp-\bp').
\end{align*}
So, we have by definition $(\tilde f_\bp,\tilde f_{\bp})=\pi^2(2\pi)^n \csch(\pi\Om)$. The normalized Bogolyubov coefficients are then related to the unnormalized ones via
\begin{align*}
\al_{k\bp}=\tilde \al_{k\bp}/\sqrt{(\tilde I_k,\tilde I_k)(\tilde f_\bp,\tilde f_\bp)},&& \be_{k\bp}=\tilde \be_{k\bp}/\sqrt{(\tilde I_k,\tilde I_k)(\tilde f_\bp,\tilde f_\bp)}.
\end{align*}

\paragraph{Bogolyubov coefficients}
Here we calculate the coefficients that take us from the Rindler plane waves to the Rindler wavepackets. The first (unnormalized) coefficient is
\begin{align*}
\tilde \al^R_{k\bp}&=(\tilde I^R_k,\tilde f^R_{\bp})=-i \integral \bx\ \tilde I_k \tilde f_\bp^*\left(ia\Om+i k-\frac{\xi}{\si ^2}\right)\nn\\
&=\mathrm{I}\times \mathrm{II},
\end{align*}
with
\begin{align*}
\mathrm{I}\equiv\integral \bx_\perp \exp\left({-\frac{\bx^2_\perp}{2\sigma_\perp^2}-i \bp_\perp\cdot \bx_\perp}\right)
\end{align*}
and
\begin{align*}
\mathrm{II}\equiv\int d\xi\ K_{-i \Om} (|\mathbf p_\perp|e^{a\xi} ) \exp{\left(-\frac{\xi ^2}{2  \si^2}+i\,\xi  k\right) }\left(a\,\Om+k+\frac{i \xi }{\si^2}\right).
\end{align*}
Integral I is a standard Gaussian integral,
\begin{align*}
\mathrm{I}=(\sqrt{2\pi}\si_\perp)^n\exp\left(-\frac{\si^2_\perp}{2}|\bp_\perp|^2\right).
\end{align*}
The Bessel function $K$ makes it impossible to solve integral II directly.  We will expand the Bessel function around $|\bp_\perp|\e^{a\xi} =0$ and then solve the integral term by term. 

 Now $|\bp_\perp|$ will be of the order of $1/\si_\perp$ due to the exponential in I. However, $\xi$ will be of order $\si$ due to the exponential in II. Therefore, our approximation only holds in the regime  where
\begin{align*}
\si_\perp\gg e^\si .
\end{align*}
From the general series representation of the Bessel function \cite{Wolfram}, 
\begin{align*}
K_{-i\Om_p}(|\mathbf p_\perp|e^{\xi} )&= \mathcal K^0+\mathcal K^1+\ldots
\end{align*}
with
\begin{align}
\mathcal K^0=\sum_{\delta'}A^{\delta'}_\xi\left(\frac{|\bp_\perp|}{2}\right)^{\delta'i\Om}, &&
\mathcal K^1= \sum_{\delta''}A^{\delta''}_{\xi'}B^{\delta''}_{\xi'}\left(\frac{|\bp_\perp|}{2}\right)^{\delta''i\Om+2},
\end{align}
where
\begin{align}
\label{eq:calcAandB}
A^{\delta'}_\xi=\half \e^{a\xi\, \delta'i \Om}\Gamma(-\delta'i\Om), &&
B^{\delta''}_{\xi'}=\frac{\e^{2a\xi'}}{1+\delta''i\Om}.
\end{align}
Here  $\delta', \delta''\in\{-1,1\}$ make complex conjugates. 

We now define II$^q$ to  represent integral II where $K$ is replaced by the $q$'th order of $K$ in $\frac{|\bp_\perp|}{2}e^{\xi}$. The resulting expectation values are defined likewise. So, for example, $\ddd^1$ is the first correction to $\ddd^0$ that is due to the expansion of $K$.

\subsubsection{Expectation values}\label{sec:calcExValDDimensionalRindlerWavepakets}
The relevant expectation values can be obtained by generalizing and solving the integrals (\ref{eq:generalExpectationValuesRinderWavepacket}).  Effectively, the  integrals only change in that $p$ is replaced by $\bp=(\Om,\bp_\perp)$, and $|p|$ by~$\Om$. 

For conciseness, we will here only show the calculation of $\ddd^0$ and $\ddd^1$. The other relevant expectation values can be obtained in a similar way and will be listed at the end of this section. 

The $D$-dimensional generalization of the expectation value of the number operator (cf. eq. \ref{eq:generalExpectationValuesRinderWavepacket}) reads 
\begin{align}
\label{eq:gereralExpectationValueOfTheNumberOperatorOfADDimensionalMode}
\ddd  &=\int_0^\infty \mathrm\!\! d\Om \integral \bp_\perp\,\frac{1}{2\sinh(\pi\Om )}\left(|\al_{k\bp}|^2e^{-\pi\Om  }+|\be_{k\bp}|^2e^{\pi\Om }\right).
\end{align}
 The effect of the expansion of Bessel function $K$ on $|\tilde\al_{k\bp}|^2$ is
\begin{align*}
|\tilde\al_{k\bp}|^2=|\mrm I|^2|\mrm{II}^0|^2+|\mrm I|^2(\mrm{II}^0\mrm{II}^{1*}+c.c.)+\ldots,
\end{align*}
where $c.c.$ stands for `the complex conjugate of the preceding term'.
The  same expression holds for  $|\tilde\be_{k\bp}|^2$ if we  make the substitution $\Om\to -\Om$,
\begin{align*}
|\tilde\be_{k\bp}|^2=|\tilde\al_{k\bp}|^2\mathlarger|_{\Om\to-\Om}.
\end{align*}
Inserting the last two equations into (\ref{eq:gereralExpectationValueOfTheNumberOperatorOfADDimensionalMode}), we have
\begin{align*}
\ddd=\ddd^0+\ddd^1,
\end{align*}
with
\begin{align}
\label{eq:calcCorrectionToN}
\ddd^1=\frac{1}{\mc N}\integral \bp \half \csch(\pi\Om)|\mathrm I|^2\left[\e^{-\pi\Om}\left(\mathrm{II}^0\mathrm{II}^{1*}+c.c.\right)+\Om\to-\Om\right],
\end{align}
where $\mc N=(\tilde g_k,\tilde g_k)(\tilde f_\bp,\tilde f_\bp)$. Here $\Om\to-\Om$ stands for `the preceding term where $\Om$ is replaced by $-\Om$'.

Let us first briefly comment on $\ddd^0$. In this order we are effectively looking at an infinitely wide Rindler wavepacket (i.e. $\si_\perp\to\infty$), which in essence should be a 1+1 dimensional Rindler wavepacket. Indeed, after performing all integrals, the expectation value of a pair of 1+1 dimensional Rindler wavepackets is reobtained (\ref{eq:calcExpectationValuesOfAPairOfRindlerWavepackets}), without any dependence on $D$. The same holds for the other relevant expectation values.

The expectation value $\langle \hat d_k^\dagger \hat d_k \rangle^1$ requires considerably more work. To keep an overview, we divide the calculation into six steps. (1) Use symmetry of the integrand to redefine the limits of integration. Isolate and solve the integral over  $\bp_\perp$. (2)  Isolate and solve the integral over $\xi$ and $\xi'$. (3) Rewrite to get a manageable form for the integral over $\Om$, and extend $\Om$ to the complex plane. Deform the contour so that the integral can be split into two parts. Write the integral as a sum of two separate contour integrals. (4) Calculate the first integral by using residues. (5) Calculate the second integral by using the saddle point method and residues. (6) Summarize all contributions to $\ddd^1$.
\begin{enumerate}
\item Observe that the integral (\ref{eq:calcCorrectionToN}) is of the form $\int_0^\infty \mathrm{d}\Om\integral \bp_\perp[f(\Om,\bp_\perp)+f(-\Om,\bp_\perp)]$, which is equal to $\int_{-\infty}^\infty \mathrm{d}\Om\integral \bp_\perp f(\Om,\bp_\perp)$. Now after some rearranging, this reads
\begin{align*}
\ddd^1=&(\sqrt{2\pi}\si_\perp)^{2n}\int_{-\infty}^\infty \!\!\mathrm{d}\Om\, \frac{ \csch(\pi\Om)}{2 \mc N} \integral \xi \mathrm d \xi'\e^{-(\xi^2+\xi'^2)/(2\si ^2)}\\
&\times \sum_{\delta}\e^{-\pi\Om}g^\delta(\xi)g^\delta(-\xi')
\left[\integral \bp_\perp \e^{-\si_\perp^2|\bp_\perp|^2}\mathcal K^0_\xi\mathcal K^1_{\xi'}\right],\nn
\end{align*}
with $g^\delta(\xi) =\e^{\delta i k \xi}\left[( a\Om+k)+\delta i \xi/\si ^2\right]$. The index $\delta\in\{-1,1\}$ turns on and off the complex conjugate from equation \ref{eq:calcCorrectionToN}. The solution of the integral over $\bp_\perp$ is 
\begin{align*}
\frac{1}{(2 \si_\perp)^2}\frac{n\pi^{n/2}}{2\si_\perp^n}\sum_{\delta',\delta''}A^{\delta'}_{\xi}A^{\delta''}_{\xi'}B^{\delta''}_{\xi'} f^{\delta'\delta''},
\end{align*}
with $A$ and $B$ as in equation \ref{eq:calcAandB} and
\begin{align*}
f^{\delta'\delta''}=(2\si_\perp)^{-(\delta'+\delta'')i\Om}\frac{\Gamma[1+\frac{n}{2}+\half(\delta'+\delta'')i\Om]}{\Gamma(1+\frac{n}{2})}.
\end{align*}
The indices $\delta',\delta''\in\{-1,1\}$ originate from the complex conjugates in $\mc K^0$ and $\mc K^1$. 


\item Integral over  $\xi$ and $\xi'$. After rearranging,
\begin{align*}
\ddd^1=&
(\sqrt{2\pi}\si_\perp)^{2n}\frac{1}{(2\si_\perp)^{2}}\frac{n\pi^{n/2}}{2\si_\perp^n}\int_{-\infty}^{\infty}\!\!\mathrm{d}\Om \,\frac{\csch(\pi\Om) }{2\mc N}
\sum_{\delta,\delta',\delta''}
f^{\delta'\delta''} \e^{-\pi\Om} \\
&\times \integral \xi' \e^{-\xi'^2/(2\si ^2)}A^{\delta''}_{\xi'}B^{\delta''}_{\xi'}g^\delta(-\xi') \\
&\times \integral \xi\,\e^{-\xi^2/(2\si ^2)}A^{\delta'}_{\xi}g^\delta(\xi).
\end{align*}
The result of the integral over $\xi$ is
\begin{align*}
\half \Gamma[-\delta'i\Om]\sqrt{2\pi}\si  \e^{\si ^2 b^2/2}\left[( a\Om+k)+\delta i b\right],
\end{align*}
with 
\begin{align}\label{eq:calcb}
b=\delta'i a\Om+\delta i k.
\end{align}
The result of the integral over $\xi'$ is
\begin{align*}
\half \frac{\Gamma(-\delta''i\Om)}{1+\delta''i\Om}\sqrt{2\pi}\e^{\si ^2b'^2/2}\si [(a\Om+k)-\delta i b'],
\end{align*}
with
\begin{align}
\label{eq:calcbPrime}
b'=\delta''i a\Om-i\delta k +2a.
\end{align}

\item Integral over $\Om$. After computing the normalization $\mc N$ and some rearranging, we have
\begin{align}
\label{eq:calcExValOfNumberOpWithMcI}
\ddd^1=&\frac{\si  n}{64 k \pi^{3/2}\si_\perp^2}\times \mathcal I,
\end{align}
where
\begin{align}
\label{eq:calcMcI}
\mathcal I=&\int_{-\infty}^{\infty}\!\!\mathrm{d} \Om\,\sum_{\delta,\delta',\delta''} \e^{\si ^2(b^2+b'^2)/2}G_\Om^{\delta\delta'\delta''},
\end{align}
with
\begin{align*}
G_\Om^{\delta\delta'\delta''}=&\,\e^{- \pi \Om}
\frac{\Gamma\left(1+\sfrac{n}{2}+\half (\delta'+\delta'')i\Om\right)}{\Gamma\left(1+\sfrac{n}{2}\right)}
\frac{\Gamma\left(-\delta'i\Om\right)\Gamma(-\delta''i\Om)}{1+\delta''i\Om}(2\si_\perp)^{-(\delta'+\delta'')i\Om}\\
&\times \left[(a \Om+k)+\delta i b\right]\left[(a \Om+k)-\delta i b'\right].
\end{align*}
The polynomial on the second line vanishes whenever 
$\delta=\delta'$.  This allows us to substitute $\delta'\to-\delta$ and drop  the sum over $\delta'$. 

After doing so, consider performing the sum in equation \ref{eq:calcMcI} that now only runs over $\delta$ and $\delta''$. There are four terms. Let $z$ be the term where $\delta=\delta''=1$. The term where $\delta=\delta''=-1$ is its complex conjugate. Thus we can write the sum of these two terms more concisely as $2\,\Re(z)$. 
Similarly, we can write the sum of the terms where $\delta=-\delta''=1$ and $\delta=-\delta''=-1$ as $2\,\Re(y)$. Thus we have after rewriting 
\begin{align*}
\mathcal I&=2\,\int_{-\infty}^{\infty}\!\!\mathrm{d} \Om\,[\Re(z)+\Re(y)],
\end{align*}
where
\begin{align*}
z&=-\exp\left\{-\si^2\Om^2+2\si^2(i+k)\Om+c\right\}\ \frac{8\pi i}{e^{2\pi\Om}-1} ,\\
y&=\exp\left\{-\si^2\Om^2+[2i\log(2\si_\perp)-2i\si^2-\pi]\Om+c\right\}\,\frac{4\,\Om }{i+\Om}\frac{\Gamma(1+\frac{n}{2}-i\Om)}{\Gamma(1+\frac{n}{2})}\Gamma^2(i\Om),\\
c&=\si^2(2-k^2-2ik).
\end{align*}
Note that $x$ and $y$ diverge separately as $\Om\to 0$ but their sum does not. To separate the integrals of $z$ and $y$, we extend $\Om$ to the complex plane and move the contour $\mc C$ up so that it runs just above the pole at $\Om=0$ instead of over the real line. Thus we have 
\begin{align*}
\mathcal I=\mathcal I_z+\mathcal I_y,
\end{align*}
with
\begin{align*}
\mathcal I_z=2\,\Re[\int_{\mathcal C}\mathrm d\Om\,z(\Om)],\qquad 
\mathcal I_y=2\,\Re[\int_{\mathcal C}\mathrm d\Om\,y(\Om)].
\end{align*}

\item $\mathcal I_z$.
Let us make the variable substitution $\Om\to \Om+i$. In this variable,
\begin{align*}
z=-\exp\left\{-\si^2[\Om^2-2k\Om+k^2-1]\right\}\frac{8\pi i}{e^{2\pi \Om}-1}.
\end{align*}
In the new variable, $\mathcal C$ runs just below the pole at $\Om=0$. Note that $z$ now satisfies $z(\Om^*)=-z^*(\Om)$.
 
Consider the contour $\mathcal C'$ that is obtained by reflecting $\mathcal C$ about the real axis. We have $\int_{\mathcal C'}\mathrm d\Om\,z(\Om)=-(\int_{\mathcal C}\mathrm d\Om\,z(\Om))^*$. By the residue theorem, 
\begin{align*}
\int_{\mathcal C}\mathrm d\Om\,z(\Om)-\int_{\mathcal C'}\mathrm d\Om\,z(\Om)=2 \pi i\, \mathrm{Res}(z,\Om=0)=8\pi \,e^{\si ^2(1-k^2)}.
\end{align*}
But the left hand side also equals
\begin{align*}
\int_{\mathcal C}\mathrm d\Om\,z(\Om)+\left(\int_{\mathcal C}\mathrm d\Om\,z(\Om)\right)^*=2\,\Re\left[\int_{\mathcal C}\mathrm d\Om\,z(\Om)\right]=\mathcal I_z.
\end{align*}
Therefore, 
\begin{align*}
\mathcal I_z=8\pi \,e^{\si^2(1-k^2)}.
\end{align*}

The total contribution to $\ddd^1$ that is caused by $z$, including the prefactors in (\ref{eq:calcExValOfNumberOpWithMcI}), is thus
\begin{equation*}
\ddd^1_z\equiv\mathcal O \left[\si e^{\si^2(1-k^2-2\omega)}\right],
\end{equation*}
where we made the definition
\begin{equation*}
\om\equiv \frac{\log(2\si_\perp)}{\si^2}.
\end{equation*}
Note that we now have two independent parameters, $\si>0$ and $\om>0$. The reason for the use of $\om$ is that it turns out to be the natural parameter in $\mathcal I_y$, as we will now see. 

\item $\mathcal I_y$. Here we use the saddle point method (see, for example, \cite{Arfken}). 
The saddle point of $y$ is at $\Om_{sp}=-i+ i\om+\mathcal O(\si^{-2})$. If we deform the contour of $\mc I_y$ so that it runs over the saddle point, the contribution to $\ddd^1$ due to this saddle point is
\begin{equation}
\langle \hat d_k^\dagger \hat d_k  \rangle^1_{y,sp}\equiv\mathcal O \left[ e^{\si^2(1-k^2-\om^2)}\right].
\end{equation}
In deforming the contour, we pick up contributions from poles that end up on the other side of the contour. The function $y$ has poles at $\Om=m i$, with $m\in\{-1,0,1,\ldots\}$. For $1<\om<2$ we do not need to cross any pole, since the contour initially runs between the pole at $\Om=0$ and $\Om=i$.

If $0<\om<1$ however, we need to cross the pole at $\Om=0$ to get to the saddle point. Likewise, for $2<\om$ we need to cross the pole at $\Om=i$, and depending on $\om$ possibly also some subsequent poles. It turns out that if $\om>1$, the pole at $\Om=i$ is always leading. 

The above results in
\begin{equation*}
\ddd^1_{y,pole}\equiv
e^{\si^2(1-k^2)}\times
\left\{\begin{array}{lll}
\mathcal O [\si e^{\si^2(1-2\om)}] &:&  0<\om <1\\
0 &:& 1<\om<2\\
\mathcal O [\si e^{4\si^2(1-\om)}] &:&  2<\om
\end{array}\right. .
\end{equation*}

\item Summarize. We are now ready to compare all contributions to $\ddd^1$. The overall result is

\begin{align*}
\ddd^1&=\ddd^1_z+\ddd^1_{y,sp}+\ddd^1_{y,pole}\\
&=e^{\si^2(1-k^2)}\times
\left\{\begin{array}{lll}
\mathcal O [\si e^{\si^2(1-2\om)}] &:&  0<\om <1\\
\mathcal O [ e^{-\si^2\om^2}] &:&  1<\om <2\\
\mathcal O [\si e^{-\si^2 2\om}] &:&  2<\omega
\end{array}\right. ,
\end{align*}
where the first line in the piecewise definition originates from the pole of $y$ at $\Om=0$, the second from the saddle point of $y$, and the third from the pole of $z$. Note that once a case has been chosen, $\om$ should be considered as a constant for the `$\mathcal O$' notation to make sense. 

Remarkably, there are two regimes, and for large $\si$ the transition between these two regimes is practically instantaneous: $\ddd^1$ is exponentially large for $\om<1-k^2/2$, but exponentially small for $\om>1-k^2/2$.
\end{enumerate}
Calculations similar to the above yield the other expectation values, the results of which are listed below. 
\paragraph{Summary of expectation values} 
Here we will give an extensive list of all relevant expectation values of a pair of $D$-dimensional Rindler wavepackets in the Minkowski vacuum. We drop the superscripts $L$ and $R$ for one-sided expectation values, since e.g. $\langle \hat d_k^{\ms R\dagger} \hat d_k^{\ms R} \rangle=\langle \hat d_k^{\ms L\dagger} \hat d_k^{\ms L}  \rangle$. To give an idea where the different pieces come from, we will write e.g. $(y\ pole)$ to denote that a term originates from a pole of the integrand $y$.  

Define  
\begin{align*}
\om=\frac{\log{2\si_\perp}}{\si^2},
\end{align*}
and let $\om_{crit}$ be the value of $\om$ that separates the regime of exponential large corrections from the regime of exponentially small corrections. Also, let
\begin{align*}
\varphi_k=\frac{\pi}{2}  \left[1+\coth \pi k \right] \left[\pi\, \coth\pi\ k - 1/k\right], 
\qquad
\vartheta_k=\frac{\pi}{2}\left[\pi  \coth ^2 \pi k -\frac{ \coth \pi  k}{k}-\frac{\pi }{2}\right].
\end{align*}
Then
\begin{align*}
\ddd&=\ddd^0+\ddd^1+\ldots\\
&=\frac{1}{e^{2\pi k}-1}\left[1+\frac{\varphi_k}{\si^2}+\mathcal O\left(\frac{1}{\si^4}\right )\right]+
e^{\si^2(1-k^2)}\times
\left\{\begin{array}{llll}
\mathcal O [\si e^{\si^2(1-2\om)}] &:&  0<\om <1 & \ms{(y\ pole)}\\
\mathcal O [ e^{-\si^2\om^2}] &:&  1<\om <2 & \ms{(y\ saddle)}\\
\mathcal O [\si e^{-\si^2 2\om}] &:&  2<\omega & \ms{(z\ pole)}
\end{array}\right. ,
\intertext{which has $\om_{crit}=1-k^2/2$,}
\langle \hat d_k^{\ms R} \hat d_k^{\ms L}\rangle&=
\frac{1}{2 \sinh \pi k}\left[1+\frac{\vartheta_k}{\si^2}+\mathcal O \left(\frac{1}{\si^4}\right)\right]
+
e^{\si^2(1-k^2)}\times
\left\{\begin{array}{llll}
\mathcal O [\si e^{\si^2(1-2\om)}] &:&  0<\om <1 & \ms{(y\ pole)}\\
\mathcal O [ e^{-\si^2\om^2}] &:&  1<\om <2 & \ms{(y\ saddle)}\\
\mathcal O [\si e^{-\si^2 2\om}] &:&  2<\omega & \ms{(z\ pole)}
\end{array}\right. ,
\intertext{which also has $\om_{crit}=1-k^2/2$,}
\langle \hat d_k \hat d_k \rangle&=\left\{\begin{array}{llll}
\mathcal O [\si^3 e^{\si^2(2-k^2-2\om)}] &:&  0<\om \leq 1-k & \ms{(y\ pole)}\\
\mathcal O [e^{\si^2(1-\om^2)}] &:&  1-k<\om \leq\sqrt{1+k^2},\ \om\neq 1 & \ms{(y\ saddle)}\\
\mathcal O [e^{-\si^2 k^2}] &:&  \sqrt{1+k^2}<\omega & \ms{\left(\langle \hat d_k \hat d_k \rangle^0\right)}
\end{array}\right., 
\intertext{which has $\om_{crit}\leq 1$, and finally,}
\langle \hat d_k^{\ms R \dagger} \hat d_k^{\ms{L}\nodagger} \rangle&=\left\{\begin{array}{llll}
\mathcal O [\si e^{\si^2(1-\om^2-4\om)}] &:&  0<\om \leq \sqrt{1+k}-1 & \ms{(y\ saddle)}\\
\mathcal O [\si e^{\si^2(1-k^2-2\om)}] &:&  \sqrt{1+k}-1<\om \leq 1/2 & \ms{(z\ pole)}\\
\mathcal O [e^{-\si^2 k^2}] &:&  1/2<\omega & \ms{\left(\langle \hat d_k^{\ms{R} \dagger} \hat d_k^\ms{L} \rangle^0\right)}
\end{array}\right.,
\intertext{where $\om_{crit}=\half(1-k^2).$}
\end{align*}
To be able to define all the functions piecewise, we had to assume $k$ to be small. 
From the expectation values above we have that, if $\om>1$ and $k$ is small, then all corrections that are due to $\si_\perp$ are exponentially small.

\acknowledgments{The authors would like to thank Patrick Hayden and Jasper van Wezel for discussion. }
\bibliographystyle{jhep}
\bibliography{bib.bib}
\end{document}